\documentclass[12pt]{iopart}
\usepackage{graphicx}

\usepackage{iopams}
\newcommand{\text}[1]{\ensuremath{\mathrm{#1}}}

\newcommand{\rvert}{|}

\usepackage[numbers, sort&compress]{natbib}
\usepackage{hyperref}
\hypersetup{colorlinks=true, linkcolor=black, citecolor=black, urlcolor=blue}

\newcommand{\topp}[1]{\ensuremath{^{({#1})}}}
\renewcommand{\vec}[1]{\ensuremath{\boldsymbol{{#1}}}}
\newcommand{\vhat}[1]{\ensuremath{\hat{\vec{{#1}}}}}

\newcommand{\phibare}{\ensuremath{\Phi}}

\newcommand{\phitot}{\ensuremath{\phibare}}

\newcommand{\green}{\ensuremath{G}}
\newcommand{\greenfree}{\ensuremath{G\topp{0}}}
\newcommand{\greensurf}{\ensuremath{G\topp{S}}}
\newcommand{\greenfsurf}{\ensuremath{\tilde{G}\topp{S}}}

\newcommand{\Cc}{\gamma}

\newcommand{\se}[1]{\ensuremath{^{\text{#1}}}}
\newcommand{\si}[1]{\ensuremath{_{\text{#1}}}}

\newcommand{\ket}[1]{\ensuremath{|#1\rangle}}
\newcommand{\me}[3]{\ensuremath{\langle#1|#2|#3\rangle}}

\newcommand{\upket}{\ket{\!\!\uparrow}}
\newcommand{\downket}{\ket{\!\!\downarrow}}

\newcommand{\dagg}{\ensuremath{^{\dagger}}}

\newcommand{\latticeA}{\ensuremath{\circ}}
\newcommand{\latticeB}{\ensuremath{\bullet}}

\newcommand{\vibc}{\ensuremath{r}}

\newcommand{\GsumA}{\mu}
\newcommand{\GsumB}{\nu}

\begin{document}
\title{Quantum simulation of the hexagonal Kitaev model with trapped ions}
\author{Roman Schmied,$^1$ Janus H. Wesenberg$^2$ and Dietrich Leibfried$^3$}
\address{$^1$ Department of Physics, University of Basel, Klingelbergstrasse 82, 4056 Basel, Switzerland}
\address{$^2$ Center for Quantum Technologies, National University of Singapore, 3 Science Drive 2, Singapore 117543, Singapore}
\address{$^3$ Time and Frequency Division, National Institute of Standards and Technology, 325 Broadway, Boulder, CO 80305, USA}
\ead{roman.schmied@unibas.ch}

\begin{abstract}
We present a detailed study of quantum simulations of coupled spin systems in surface-electrode ion-trap arrays, and illustrate our findings with a proposed implementation of the hexagonal Kitaev model [A.~Kitaev, Annals of Physics \textbf{321},2 (2006)]. The effective (pseudo)spin interactions making up such quantum simulators are found to be proportional to the dipole--dipole interaction between the trapped ions, and are mediated by motion which can be driven by state-dependent forces. The precise forms of the trapping potentials and the interactions are derived in the presence of a surface electrode and a cover electrode. These results are the starting point to derive an optimized surface-electrode geometry for trapping ions in the desired honeycomb lattice of Kitaev's model, where we design the dipole--dipole interactions in a way that allows for coupling all three bond types of the model simultaneously, without the need for time discretization. Finally we propose a simple wire structure that can be incorporated in a microfabricated chip to generate localized state-dependent forces which drive the couplings prescribed by this particular model; such a wire structure should be adaptable to many other situations.
\end{abstract}

\section{Introduction}

Ion trap systems have proven to perform well for implementing the basic elements of traditional quantum computing, where evolution is described in terms of discrete gate operations, which can be implemented step by step as intermediate states are irrelevant. This is in contrast to quantum simulations, where the goal is to simulate the \emph{continuous} evolution of a given Hamiltonian.
While the initial proposal for quantum computing with trapped ions relied on a number of sequential steps to mediate effective qubit interactions \cite{Cirac1995}, other approaches \cite{Soerensen1999,Moelmer1999a,Mintert2001,Leibfried2003b,Porras2004,GarciaRipoll2005,Porras2006b,Bermudez2009,Welzel2011} achieve interaction between the internal states of the ions via  constant Hamiltonians and therefore allow the development of quantum simulators based on trapped ions \cite{Wunderlich2002,Deng2005,GarciaRipoll2005,Porras2006b,Friedenauer2008,Johanning2009,Kim2010}. In such simulators, interactions between trapped ions are dominated by the Coulomb potential. For this interaction to affect internal states (i.e., the qubits or pseudo-spins representing the effective quantum system to be simulated), state-dependent forces must be applied to some or all of the trapped ions. State-dependent forces can be achieved through optical ac Stark shifts~\cite{soerensen00:entan_quant_comput_with_ions,milburn00:ion_trap_quant_comput_with,Leibfried2003b,GarciaRipoll2005,Lee2005,Kim2009,Kim2010}, static magnetic-field gradients in combination with homogeneous radio-frequency (rf) fields~\cite{Mintert2001,Johanning2009,Johanning2009b,Welzel2011}, or with rf field gradients~\cite{Ospelkaus2008,Ospelkaus2011}. While in most cases the Coulomb interaction is considered between ions in a self-assembled single chain or crystal, coupling of independently trapped ions has recently been demonstrated~\cite{Brown11:Coupled-quantiz,Harlander11:Trapped-ion-ant}.

For quantum simulations with ions in microtraps, we must take into account how the presence of the electrodes modifies the Coulomb interaction. While in many systems this effect is negligible (for example, in the surface-electrode setup of~\cite{Brown11:Coupled-quantiz} the Coulomb coupling was found to be enhanced by only $1.8\,\%$, in agreement with our more general results in \sref{sec:dipdip}), the general theory developed here for a lattice of surface-electrode (SE) microtraps shows that significant modifications to free-space couplings are possible. Far from being an inconvenience, these modified interactions can be used to design quantum simulations with specific short-range effective pseudo-spin interactions, which we illustrate with the hexagonal Kitaev model as a concrete example.

The remainder of this paper is organized as follows. In \sref{sec:cover-plane-boundary} we present a Green's function approach to solving electrostatic problems as they occur for surface-electrode ion traps in the presence of a cover electrode. In \sref{sec:spinspin} we derive general expressions for spin--spin couplings in two-dimensional microtrap arrays, applicable, for example, to electric coupling to light fields or magnetic coupling to microwave near-field gradients. In \sref{sec:Kitaev} we combine all these methods to show how the hexagonal Kitaev model~\cite{Kitaev2006}  can be implemented with an array of trapped ions on an optimized surface-electrode chip, including a dedicated wire structure that could be integrated in the chip to simultaneously mediate the couplings along three distinct bonds by use of magnetic-field gradients. Finally, \ref{app:coordinates} gives a summary of the used coordinate systems.

\section{Electrostatics in the presence of conducting planes}
\label{sec:cover-plane-boundary}

\begin{figure}
	\centering
	\includegraphics[width=0.5\linewidth]{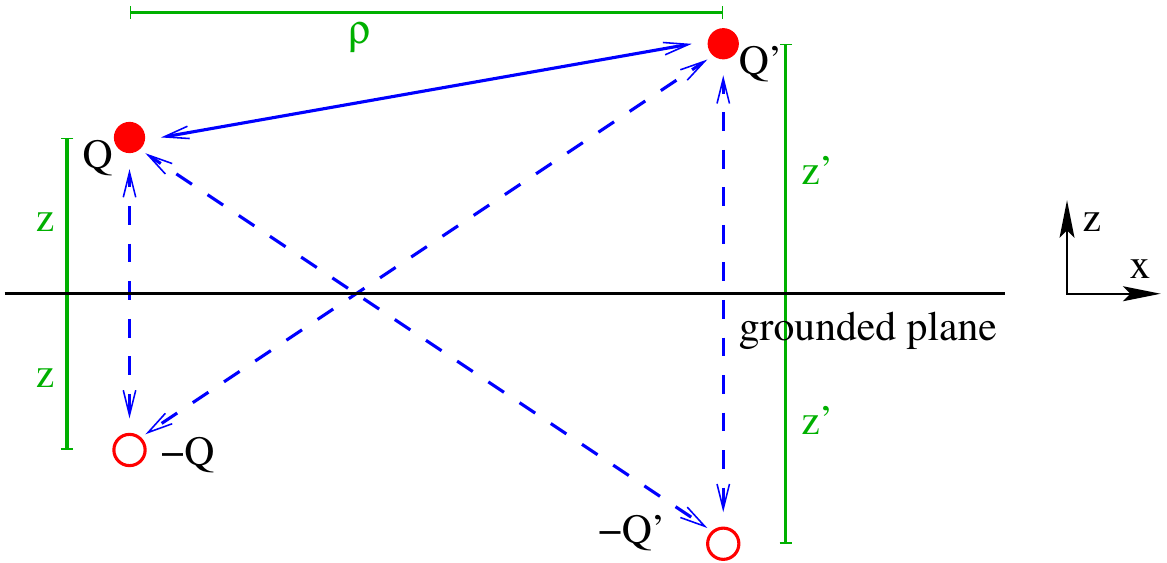}
	\caption{Coulomb interaction between two charges $Q$ and $Q'$ (full red circles) in the presence of a grounded plane. The image charges (empty red circles) are located below the grounded plane and carry opposite charge. Interactions between the charges (full blue arrow) contribute to~\eref{eq:Coulomb} in full, while interactions between charges and image charges (dashed blue arrows) contribute with a prefactor of $1/2$~\cite{Taddei2009}.}
	\label{fig:interaction}
\end{figure}

The electrostatic interaction between charged particles close to conducting surfaces can be strongly modified by the presence of the conductors~\cite{Jackson}. In the idealized geometry of a perfectly conducting grounded electrode plane at $z=0$, the total electrostatic energy of a set of charges $Q_i$ located at positions $\vec{r}_i$ in the half-space $z_i>0$ (see \fref{fig:interaction}) is~\cite{Taddei2009}
\begin{equation}
    \label{eq:Coulomb}
    E_{\infty}\se{C} = \frac{1}{4\pi\epsilon_0} \left[
        -\sum_i \frac{Q_i^2}{4z_i}
        + \sum_{i<j} Q_i Q_j G_{\infty}(\vec{r}_i,\vec{r}_j)
    \right].
\end{equation}
The Coulomb interaction term in~\eref{eq:Coulomb} is expressed in terms of the Dirichlet Green's function $\green_{\infty}$, which can be found from the free-space Green's function $\greenfree(\vec{r},\vec{r}') = 1/\|\vec{r}-\vec{r}' \|$ by the method of images (see \fref{fig:interaction}),
\begin{equation}
  \label{eq:green}
  \green_{\infty} (\vec{r},\vec{r}')
  = \frac{1}{\sqrt{\rho^2 +(z-z')^2}}-\frac{1}{\sqrt{\rho^2 +(z+z')^2}},
\end{equation}
where $\rho=\sqrt{(x-x')^2+(y-y')^2}$ is the horizontal distance between the charges. 

In the following we review the effects of a grounded cover plane, \emph{i.e.}, a second, parallel conducting plane covering the electrode plane at height $z=H$ (see \fref{fig:cpregimes}a).  In the initial proposal \cite{Chiaverini2005} and demonstration \cite{Seidelin2006} of surface-electrode rf traps, the conducting surface nearest to the trap electrodes was theoretically at infinity but in practice a part of the surrounding apparatus. It has been suggested that adding a cover plane in the form of a dc-biased mesh above the electrodes could improve trap depth \cite{Pearson2006}. In addition to possible benefits of providing bias field and shielding, the cover plane could have more practical advantages, namely shielding the trapping region from fields due to quasi-static charges on insulators in the vacuum chamber, and establishing a more well-defined boundary condition. Further, if the cover plane is modified to carry rf and dc electrodes of arbitrary shape in the same way as the electrode plane, the presented formulas can be used directly to calculate the combined electric fields generated in this ``sandwich trap'' geometry (however, if optical access to such a trap geometry is achieved with holes and/or fiber optics in the electrode planes~\cite{VanDevender2010}, the present full-plane treatment must be adapted~\cite{Schmied2010}).

Below, we first modify the Green's function~(\ref{eq:green}) to include the cover plane and illustrate that a cover plane at height $H$ leads to exponential shielding on a lateral length scale of $H$ (\sref{sec:screen-effects-cover-plane}), then consider its effects on the electric potential generated by surface electrodes (\sref{sec:surf-greens-funct}) and on effective dipole--dipole interactions between vibrating trapped ions (\sref{sec:dipdip}).

\subsection{the shielding effect of the cover plane}
\label{sec:screen-effects-cover-plane}

When a grounded conducting cover plane at height $z=H$ is added to the setup of \fref{fig:interaction}, the Coulomb interaction~\eref{eq:Coulomb} of charges located between these two planes is modified to
\begin{equation}
	\label{eq:CoulombE}
	E\si{C} = \frac{1}{4\pi\epsilon_0} \left[
		-\sum_i Q_i^2 e_H(z_i)
		+ \sum_{i<j} Q_i Q_j G_H(\vec{r}_i,\vec{r}_j) \right].
\end{equation}
Both the scaled self-potential $e_H(z)$ and the Dirichlet Green's function $\green_H$  corresponding to the cover plane geometry with infinite conducting electrode planes at $z=0$ and $z=H$ can be found by summing over an infinite sequence of mirror planes; and in the absence of a cover plane ($H\to\infty$) they reduce to~\eref{eq:Coulomb} and~\eref{eq:green}. The scaled self-potential is $e_H(z)=-[2\gamma+\psi(z/H)+\psi(1-z/H)]/(4H)=\frac{1}{4z}+\mathcal{O}(z^2/H^3)$ in terms of Euler's constant $\gamma=0.577216\ldots$ and the digamma function $\psi(a)=\Gamma'(a)/\Gamma(a)$. The Dirichlet Green's function is
\numparts
\begin{eqnarray}
  \green_H(\vec{r},\vec{r}')
  \label{eq:greencp}
  \label{eq:greencpa}
 &=\sum_{\GsumA=-\infty}^\infty \green_\infty(\vec{r}+2 \GsumA H \vhat{z} ,\vec{r}')\\
  \label{eq:greencpb}
  &= \sum_{\GsumB=1}^\infty \frac{4}{H} \sin\left(\frac{\GsumB \pi  z}{H}\right)  \sin\left(\frac{\GsumB \pi  z'}{H}\right)
  K_0\left( \frac{\GsumB \pi \rho}{H}\right),
\end{eqnarray}
\endnumparts
where $K_0$ is the modified Bessel function of the second kind. The second form~\eref{eq:greencpb} is obtained by solving the Laplace equation in cylindrical coordinates~\cite{Jackson}. Both forms converge for all parameters $(\rho,z,z')$; but while \eref{eq:greencpa} converges faster when $\|\vec{r}-\vec{r}'\|\lesssim H$, \eref{eq:greencpb} is more suitable if $\|\vec{r}-\vec{r}'\|\gtrsim H$, in particular for $\rho\gg H$, as discussed below.

\begin{figure}
	\centering
	\includegraphics[width=\linewidth]{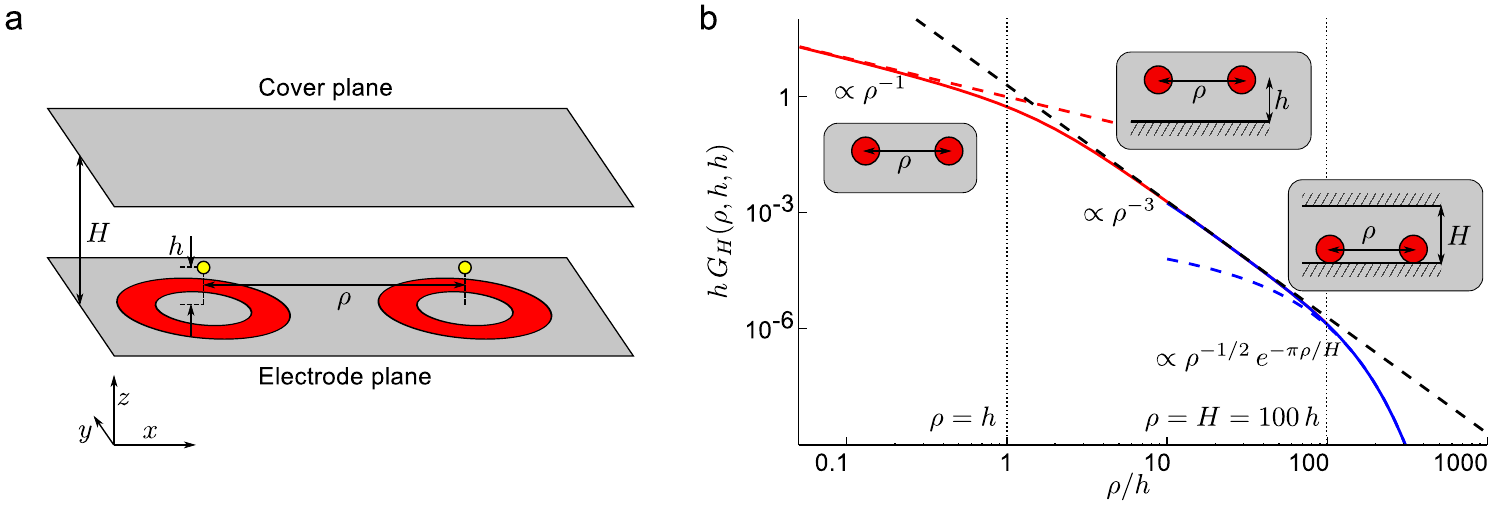}
	\caption{
  		\textbf{(a)} Sketch of a surface-electrode trap with a grounded cover plane positioned at a height $H$ above the electrode plane. The red ring electrodes are at rf potential, while all grey areas are grounded. For static interactions or interactions varying slowly compared to the rf period, only the time-averaged potential contributes, so for our purposes the situation is equivalent to two completely grounded planes.
		\textbf{(b)} The interaction energy (\ref{eq:CoulombE}) between two point charges at same height $h$ over the electrode plane, as a function of the charge separation $\rho$, in the presence of a cover plane at height $H=100 h$. The red and blue parts of the solid curves are computed by~\eref{eq:greencpa} and~\eref{eq:greencpb}, respectively, while the dashed lines illustrate the approximate behavior given by~(\ref{eq:screen_asympt}).}
	\label{fig:cpregimes}
\end{figure}

The Coulomb interaction energy $\green_H(\rho,z,z') Q Q'/(4\pi \epsilon_0)$ between two charged particles in a SE trap depends on the horizontal separation $\rho$, as illustrated in \fref{fig:cpregimes}b. To illustrate this interaction energy we take the particles to be at the same height $h$ above the electrode plane, and the cover plane height $H$ to be much larger than $h$.
When $\rho\ll H$, we expect the cover plane to be irrelevant, so that the interaction is described by a single image charge: it falls of as $\rho^{-1}$ while $\rho < h$ (where the electrode plane is irrelevant) and as $\rho^{-3}$ thereafter, as described by~\eref{eq:green}.
When $\rho \gtrsim H$ the cover plane becomes important and the asymptotically dominant form is the first  term of the resummation~\eref{eq:greencpb}, so that the presence of the cover plane leads to an exponential shielding at the length scale of the cover plane height, as illustrated in Fig.~\ref{fig:cpregimes}. Summarizing,
\begin{equation}
  \label{eq:screen_asympt}
  \green_H(\rho, h,h) \approx
  \cases{
    {1/\rho} & for $\rho\ll h$\\
    2 {h^2/\rho^3} & for $h\ll \rho \ll H$\\
    \sqrt{\frac{8 }{H\rho}} \sin^2\left(\frac{\pi h}{H}\right)  e^{- \pi \rho/H} & for $\rho\gg H$.
   }
\end{equation}

\subsection{the potential due to the surface electrodes}
\label{sec:surf-greens-funct}

The contribution to the total potential from the structured electrodes in the $z=0$ plane can be computed as an integral over the electrode plane:
\begin{equation}
  \label{eq:greenssurffold}
  \phitot(\vec{r}) = \int_{z'=0} \greensurf_H(\vec{r},\vec{r}') \phitot(\vec{r}') \rmd x'\rmd y',
\end{equation}
where we have introduced a ``surface Green's  function''
$\greensurf_H(\vec{r},\vec{r}')  \equiv \left. \frac{1}{4\pi} \frac{\partial}{\partial z'} \green_H(\vec{r},\vec{r}') \right\rvert_{z'=0}$.
In the absence of a cover plane, the surface Green's function was found to be~\cite{Schmied2010}
\begin{equation}
\label{eq:greensurf}
\greensurf_\infty(\vec{r},\vec{r}') = \greensurf_\infty(\rho, z) =\frac{z}{2\pi \left(\rho^2+z^2\right)^{3/2}},
\end{equation}
with the geometric interpretation that the potential at $\vec{r}$ due to an electrode at potential $\Phi_0$ is $\Phi_0/2 \pi$ times the solid angle spanned by the electrode as seen from $\vec{r}$ \cite{Oliveira2001}. Alternatively, the electric field at $\vec{r}$ is proportional to the magnetic field that would be observed if a current were flowing along the edge of the electrode \cite{Oliveira2001,Schmied2010b}. For electrode configurations that are translationally invariant in the $x$-direction, the system can be described by conformally mapping the upper-half $y z$-plane ($z>0$) to a disc \cite{Wesenberg2008}.
Analogous to~\eref{eq:greencpa} and~\eref{eq:greencpb}, we have for the general case including a cover plane,
\numparts
\begin{eqnarray}
	\label{eq:greencpsa}
	\greensurf_H(\rho,z)
	&=\sum_{\GsumA=-\infty}^\infty \greensurf_\infty(\rho, z+2 \GsumA H)\\
	\label{eq:greencpsb}
	&=\frac{1}{H^2} \sum_{\GsumB=1}^\infty \GsumB \sin\left(\frac{\GsumB \pi  z}{H}\right)   K_0\left( \frac{\GsumB \pi \rho}{H} \right)\\
	\label{eq:greencpsc}
	&=\greensurf_{\infty}(\rho, z) - \frac{1}{4 \pi H^2}\sum_{j=1,3,\ldots}^\infty (j+1)\zeta(j+2) \left(\frac{s}{2 H}\right)^j  P_j\left(\frac{z}{s}\right),
\end{eqnarray}
\endnumparts
where $\zeta$ is the Riemann zeta function, $P_j$ are Legendre polynomials, and $s=\|\vec{r}-\vec{r}'\|=\sqrt{\rho^2+z^2}$. Forms~\eref{eq:greencpsa} and~\eref{eq:greencpsb} converge for all $(\rho,z)$; but while \eref{eq:greencpsa} converges faster when $s\lesssim H$, \eref{eq:greencpsb} is more suitable if $s\gtrsim H$. Form~\eref{eq:greencpsc} is restricted to $s<2H$ and is most useful for $s\ll H$. Similar to~\eref{eq:screen_asympt} we find the approximate behaviors
\begin{equation}
	\label{eq:screen_asympt_s}
	\greensurf_H(\rho, z) \approx
	\cases{
		\frac{\psi'(\frac{z}{2H})-\psi'(1-\frac{z}{2H})}{8\pi H^2} & for $\rho\ll z$\\
		\frac{1}{\sqrt{2\rho H^3}}\sin\left(\frac{\pi z}{H}\right)e^{-\pi\rho/H} & for $\rho\gg H$,
	}
\end{equation}
where $\psi'(a)=\Gamma''(a)/\Gamma(a)-\psi^2(a)$ is the first derivative of the digamma function. We conclude that the influence of any surface electrode is exponentially damped at distances larger than $H$, which is advantageous for the experimental construction of quasi-infinite surface microtrap lattices in that it reduces the influence of the inevitable electrode boundary: at any point further than $H$ away from the edge of the electrode and cover plane, the trap will look as if the electrode were infinitely large.

Since the surface Green's function only depends on the $x$ and $y$ coordinates through $\vec{r}-\vec{r}'$, \eref{eq:greenssurffold} is a folding integral (convolution)~\cite{Schmied2010} and can be rewritten as a product of the Fourier-transformed quantities, $\tilde{\phitot}(k_x,k_y,z) =\greenfsurf_H(k_x,k_y,z) \tilde{\phitot}(k_x,k_y,0)$, with
\begin{equation}
	\tilde{\phitot}(k_x,k_y,z) = \frac{1}{2\pi}\int_{-\infty}^{\infty} \phitot(x,y,z) e^{-\rmi(k_xx+k_yy)}\rmd x\,\rmd y
\end{equation}
and a similar expression for the Fourier-transformed Green's function. The latter is cylindrically symmetric ($k=\sqrt{k_x^2+k_y^2}$),
\begin{equation}
	\label{eq:greenfsurf}
	\greenfsurf_H(k,z) = \frac{\sinh(k H - k z)}{\sinh(k H)}
	\to e^{-k z}\text{\ for\  }H\to\infty,
\end{equation}
and allows a rather intuitive interpretation. All solutions of the Laplace equation with horizontal wavevector $\{k_x,k_y\}$ are of the form $e^{\rmi(k_x x+k_y y)}(\alpha_+e^{+k z}+\alpha_-e^{-k z})$; the Green's function~\eref{eq:greenfsurf} gives the unique solution which has unit amplitude on the electrode plane [$\greenfsurf_H(k,0)=1$] and zero amplitude on the cover plane [$\greenfsurf_H(k,H)=0$]. Therefore~\eref{eq:greenfsurf} gives the unique extension of a unit-amplitude potential plane wave from the $z=0$ plane into the $z>0$ half-space which satisfies the boundary condition of vanishing amplitude on the cover plane. The fact that the momentum-space representation of the surface Green's function~\eref{eq:greenfsurf} can be written without infinite sums greatly simplifies the description of infinite lattices of surface-electrode microtraps~\cite{Schmied2009}.

\subsection{dipole--dipole interactions between trapped ions}
\label{sec:dipdip}

Trapped-ion quantum simulators couple internal degrees of freedom of the ions (typically hyperfine states or metastable D-states) through a state-dependent coupling to shared vibrational degrees of freedom~\cite{Cirac1995,Porras2004,Porras2006b,Ospelkaus2008,Johanning2009,Johanning2009b} (see \sref{sec:spinspin}). A crucial ingredient of these couplings is the precise nature of the Coulomb interactions between the ions. Here we address the details of this latter point, since it will determine how to \emph{construct} a quantum simulator of a desired system, as exemplified in \sref{sec:Kitaev}.

We consider the regime of ``stiff'' ion trapping~\cite{Porras2004}, where the Coulomb interaction is relatively small compared to the trapping potential, and we can interpret the normal-mode dynamics of the ion crystal as that of a set of \emph{local} harmonic oscillators that are weakly coupled. The ion trapping potential defines a set of local eigenmodes for the $i\se{th}$ ion corresponding to vibration in three orthogonal directions $\vec{m}_i^{\mu}$ (with $\|\vec{m}_i^{\mu}\|=1$ for $\mu=1,2,3$) around an equilibrium position $\vec{R}_{0i}$. In what follows we use these directions to parametrize the position of the $i\se{th}$ ion as
\begin{equation}
	\label{eq:localcoord}
	\vec{r}_i=\vec{R}_{0i}+\sum_{\mu=1}^3\vibc_i^{\mu}\vec{m}_i^{\mu}.
\end{equation}

The total Coulomb energy of a set of $N$ charges is given in~\eref{eq:CoulombE}, and the leading-order term that couples the motion of the ions is
 \begin{equation}
	\label{eq:ddcoupling}
	E\si{C}\se{coupling} = \frac{1}{4\pi\epsilon_0} \sum_{i<j}^N \sum_{\mu,\nu=1}^3 Q_i Q_j \vibc_i^{\mu} \vibc_j^{\nu} \vec{m}_i^{\mu} \cdot \vec{\nabla}_i\vec{\nabla}_j G_H(\vec{R}_{0i},\vec{R}_{0j})\cdot\vec{m}_j^{\nu}.
\end{equation}
Since we are mainly interested in near(est)-neighbor interactions, we evaluate this expression in terms of the infinite sum over image charge pairs~\eref{eq:greencp}, rather than the resummed form~\eref{eq:greencpb}:
\begin{equation}
	\label{eq:dipoleH}
	\vec{m}\cdot\vec{\nabla}\vec{\nabla'}G_H(\vec{r},\vec{r'})\cdot\vec{m'} =
	\sum_{\GsumA=-\infty}^{\infty} \vec{m}\cdot\vec{\nabla}\vec{\nabla'}G_{\infty}(\vec{r}+2\GsumA H\vec{\hat{z}},\vec{r'}) \cdot\vec{m'},
\end{equation}
where the explicit dipole--dipole coupling is given by the expression without a cover plane,
\begin{equation}
	\label{eq:dipdip1}
\fl	\vec{m}\cdot\vec{\nabla}\vec{\nabla'}G_{\infty}(\vec{r},\vec{r'})\cdot\vec{m'} =
	\frac{\vec{m}\cdot\vec{m'}-3(\vec{m}\cdot\vec{n})(\vec{m'}\cdot\vec{n})}{\|\vec{r}-\vec{r'}\|^3}
	 -\frac{\vec{m}\cdot\vec{\bar{m}'}-3(\vec{m}\cdot\vec{\bar{n}})(\vec{\bar{m}'}\cdot\vec{\bar{n}})}{\|\vec{r}-\vec{\bar{r}'}\|^3}
\end{equation}
in terms of $\vec{n}=(\vec{r}-\vec{r'})/\|\vec{r}-\vec{r'}\|$, $\vec{\bar{n}}=(\vec{r}-\vec{\bar{r}'})/\|\vec{r}-\vec{\bar{r}'}\|$, and the mirrored quantities $\vec{\bar{r}'}=\vec{r'}-2(\vec{r'}\cdot\vec{\hat{z}})\vec{\hat{z}}$ and $\vec{\bar{m}}'=\vec{m'}-2(\vec{m'}\cdot\vec{\hat{z}})\vec{\hat{z}}$. The first term of~\eref{eq:dipdip1} is the well-known dipole--dipole interaction, while the second term is the correction due to image charges in the electrode.

In order to illustrate the behavior of the dipolar interaction~\eref{eq:dipdip1} in close proximity of a conducting electrode plane, we again consider two ions located at equal height $h$ above the electrode plane, spaced by a distance $\rho$ along the $x$ axis, and in the absence of a cover plane. If we assume that both ions vibrate along axes $\vec{m}=\vec{m'}$ that are parallel to the lab-frame coordinate axes, then we find that the presence of the electrode plane can either increase or decrease the dipolar coupling strength:
\begin{eqnarray}
	\label{eq:dipdipdemo}
\fl	\vec{m}=\vec{m'}=\vec{\hat{x}}:\quad &
	\vec{\hat{x}}\cdot\vec{\nabla}\vec{\nabla'}G_{\infty}(h\vec{\hat{z}},h\vec{\hat{z}}+\rho\vec{\hat{x}})\cdot\vec{\hat{x}}
	= -\frac{2}{\rho^3} \left[ 1+\frac{\rho^3(2h^2-\rho^2)}{(\rho^2+4h^2)^{5/2}}\right]
	\stackrel{\rho\gg 2h}{\longrightarrow} -\frac{24h^2}{\rho^5}\nonumber\\
\fl	\vec{m}=\vec{m'}=\vec{\hat{y}}: &
	\vec{\hat{y}}\cdot\vec{\nabla}\vec{\nabla'}G_{\infty}(h\vec{\hat{z}},h\vec{\hat{z}}+\rho\vec{\hat{x}})\cdot\vec{\hat{y}}
	= +\frac{1}{\rho^3} \left[ 1-\frac{\rho^3}{(\rho^2+4h^2)^{3/2}}\right]
	\stackrel{\rho\gg 2h}{\longrightarrow} +\frac{6h^2}{\rho^5}\nonumber\\
\fl	\vec{m}=\vec{m'}=\vec{\hat{z}}: &
	\vec{\hat{z}}\cdot\vec{\nabla}\vec{\nabla'}G_{\infty}(h\vec{\hat{z}},h\vec{\hat{z}}+\rho\vec{\hat{x}})\cdot\vec{\hat{z}}
	= +\frac{1}{\rho^3} \left[ 1-\frac{\rho^3(8h^2-\rho^2)}{(\rho^2+4h^2)^{5/2}}\right]
	\stackrel{\rho\gg 2h}{\longrightarrow} +\frac{2}{\rho^3}.
\end{eqnarray}
Thus we see that by choosing the directions of vibration $\vec{m}$ in particular ways we can use the presence of the electrode plane to make the dipolar interactions fall off with the fifth power of distance instead of with the third power (which is the case in the absence of any conducting planes), as long as the ion oscillation frequency is low enough to avoid the effects of retardation and dissipation. The relevant length scale that determines whether or not the electrode plane has a strong influence on the dipole--dipole coupling is $\rho\sim2h$, similar to \fref{fig:cpregimes}; for even farther separations ($\rho>H$) we find exponentially damped dipole--dipole couplings due to the shielding effect of the cover plane (see \sref{sec:screen-effects-cover-plane}). These rapid dampings can be used to construct lattice simulation models with nearly local interactions, which is a desirable feature since many spin models from condensed-matter physics are formulated in terms of such local (\emph{e.g.}, nearest-neighbor) couplings.

\section{Spin--spin interactions between trapped ions}
\label{sec:spinspin}

This section derives how state-dependent forces can induce pseudo-spin interactions between neighboring ions through the Coulomb potential. While this effect is well-known in principle~\cite{Cirac1995}, we show how these effective interactions are constructed in a \emph{lattice} of ions without the need for time-slicing (``Trotterization''~\cite{Lloyd1996}). Further we show that, to lowest order, the effective interaction strengths are proportional to the real-space Coulomb coupling strengths, an observation that greatly simplifies the design of lattice-based quantum simulators (see \sref{sec:Kitaev}).

\subsection{normal modes of vibration}
\label{sec:normalmodes}

For small oscillation amplitudes $\vibc_i^{\mu}$ the coupled harmonic motion of $N$ ions in a lattice can be described by considering the local trapping potential curvatures (the second derivatives of the ion trapping pseudopotential with respect to position) around the equilibrium positions $\vec{R}_{0i}$ and including the Coulomb couplings between ions in separate wells to second order~\cite{gol80}. If we assume that all excursions $\vec{r}_i-\vec{R}_{0 i}$ are already written in terms of the ``bare'' eigenmodes of the isolated local trapping potentials (with ``bare'' frequencies $\bar{\omega}_{i\mu}$), as in~\eref{eq:localcoord}, then the potential energy of the ions of mass $M$ is
\begin{equation}
	\label{eq:pot2}
	V = \frac12 M \left[ \sum_{i=1}^N \sum_{\mu=1}^3 \bar{\omega}_{i\mu}^2 (\vibc_i^{\mu})^2
	+ \sum_{i,j=1}^N \sum_{\mu,\nu=1}^3 \Cc^{\mu\nu}_{ij} \vibc_i^{\mu}\vibc_j^{\nu} \right],
\end{equation}
where $\Cc^{\mu\nu}_{ii}=0$ and $\Cc^{\mu\nu}_{i\neq j} = \frac{Q_i Q_j}{4\pi\epsilon_0 M} \vec{m}_i^{\mu} \cdot \vec{\nabla}_i\vec{\nabla}_j G_H(\vec{R}_{0i},\vec{R}_{0j})\cdot\vec{m}_j^{\nu}$, see~\eref{eq:ddcoupling}. This quadratic potential energy can be diagonalized using coefficients $O_{i\mu m}$ such that the real-space displacements can be written as
\begin{equation}
	\label{eq:normalmodes}
\fl	\vibc_i^{\mu} = \sum_{m=1}^{3N} O_{i\mu m} q_m
	\qquad \text{with}\ \sum_{m=1}^{3N} O_{i\mu m}O_{j\nu m}=\delta_{i j}\delta_{\mu\nu}
	\text{\ and}\ \sum_{i=1}^{N} \sum_{\mu=1}^3 O_{i\mu m}O_{i\mu m'}=\delta_{m m'},
\end{equation}
and $V = \sum_{m=1}^{3N} \frac12 M \omega_m^2 q_m^2$ in terms of the lattice normal-mode amplitudes $q_m$ and their frequencies $\omega_m$. We quantize these normal modes through $q_m\mapsto\hat{q}_m=q_{0m}(\hat{a}_m+\hat{a}\dagg_m)$ with $q_{0 m}=\sqrt{\hbar/(2M\omega_m)}$ and the usual commutation relations $[\hat{a}_m,\hat{a}\dagg_{m'}]=\delta_{m m'}$. We will work in the ``stiff'' \label{stiff} lattice limit, where we assume that the ``bare'' trap frequencies $\bar{\omega}_{i\mu}\equiv\bar{\omega}_{\mu}$ $\forall i$ are equal for all ions\footnote{This equality can be relaxed to the condition that for the modes that are used for inducing spin--spin couplings, the spreads of the bare frequencies are much smaller than the dominant Coulomb couplings.} and the Coulomb couplings between ions do not significantly mix local modes with different values $\mu$. This is the case when (i) the bare trap frequencies are sufficiently far apart: $|\Cc_{i j}^{\mu\nu}|\ll|\bar{\omega}_{\mu}^2-\bar{\omega}_{\nu}^2|$ $\forall i,j,\mu\neq\nu$, and (ii) the vibrational bands are sufficiently narrow: $|\Cc_{i j}^{\mu\mu}|\ll\min_{\nu}|\bar{\omega}_{\mu}^2-\bar{\omega}_{\nu}^2|$ $\forall i,j,\mu$. In this limit, the normal modes of the lattice separate into three disjoint sets indexed by $\mu\in\{1,2,3\}$, each containing $N$ normal modes with frequencies close to the corresponding $\bar{\omega}_{\mu}$.

\subsection{state-dependent forces}

In addition to the Coulomb couplings, which are always on and define the coupled vibrational eigenmodes of the trapped ions, we can experimentally introduce fields that couple to \emph{internal} states of the ions. Examples of such interactions are electric or magnetic dipole couplings, Raman couplings, or electric quadrupole couplings to laser or microwave fields. In the following general treatment we assume that there is a coupling between (a) classical external field(s) effectively oscillating with angular frequency $\omega\si{I}$, and two internal states of each ion, forming an effective two-level (spin-1/2 or pseudo-spin-1/2) system.
Irrespective of the type of induced coupling, the coupling operator of the $i\se{th}$ ion in its (pseudo)spin-1/2 subspace can be expressed as a linear combination of the identity operator $\hat{\sigma}_0^{(i)}$ and the Pauli matrices $\hat{\sigma}_{\ell}^{(i)}$, $\ell\in\{X,Y,Z\}$ expressed in a quantization coordinate frame whose axes are given by the orthonormal vectors $\vec{\hat{X}}$, $\vec{\hat{Y}}$, $\vec{\hat{Z}}$ (see~\ref{app:coordinates}). The coupling Hamiltonian can thus be very generally expressed as
\begin{equation}
	\label{eq:Hint}
	\mathcal{H}\si{I}
	\approx \sum_{i=1}^N\sum_{\ell\in\{0,X,Y,Z\}} [c_{\ell}\topp{i} \cos(\omega\si{I}t+\phi\topp{i}\si{c})
		+(\vec{r}_i-\vec{R}_{0i})\cdot\vec{s}_{\ell}\topp{i} \cos(\omega\si{I}t+\phi\topp{i}\si{s})]\hat{\sigma}_{\ell}\topp{i},
\end{equation}
where we have performed a  first-order expansion in the ion positions assuming small oscillation amplitudes. Any type of spin-1/2 coupling that is used with trapped ions (including effective couplings to pseudo-spin degrees of freedom) can be brought into this form, where terms with non-vanishing prefactors $c_{\ell}^{(i)}$ and $\vec{s}_{\ell}^{(i)}$  are referred to as ``carrier'' and ``sideband'' terms, respectively. The phases can absorb differences in the details of driving fields: while stationary fields in general have $\phi\si{c}\topp{i}=\phi\si{s}\topp{i}$, travelling waves (\emph{e.g.}, light fields) are characterized by $\phi\si{c}\topp{i}=\phi\si{s}\topp{i}\pm\frac{\pi}{2}$.

As an example, the coupling of \emph{physical} spins to a magnetic field is found by expanding their magnetic dipole operators as $\label{dipoleop}\vec{\hat{\mu}}^{(i)}=g\topp{i}\mu\si{B}(\hat{\sigma}_X\topp{i}\vec{\hat{X}}+\hat{\sigma}_Y\topp{i}\vec{\hat{Y}}+\hat{\sigma}_Z\topp{i}\vec{\hat{Z}})$ in terms of the Bohr magneton $\mu\si{B}$ and the $g$-factors $g\topp{i}$; for small ion excursions the coupling Hamiltonian to the magnetic field $\mathcal{H}\si{I}=-\sum_i\vec{\hat{\mu}}\topp{i}\cdot\vec{B}(\vec{r}_i)\cos(\omega\si{I}t+\phi)$ can thus be expressed in the form of~\eref{eq:Hint} with
\begin{eqnarray}
	\label{eq:magneticSC}
	c_{\ell}\topp{i} &= \cases{
		0\rule{4cm}{0pt} & for $\ell=0$\\
		-g\topp{i}\mu\si{B}\vec{\hat{X}}\cdot\vec{B}(\vec{R}_{0i}) & for $\ell=X$, and similarly for $\ell=Y,Z$
	}\nonumber\\
	\vec{s}_{\ell}^{(i)} &= \cases{
		0\rule{4cm}{0pt} & for $\ell=0$\\
		-g\topp{i}\mu\si{B}\vec{\nabla} [\vec{\hat{X}}\cdot\vec{B}(\vec{R}_{0 i})] & for $\ell=X$, and similarly for $\ell=Y,Z$
	}
\end{eqnarray}
and $\phi\topp{i}\si{c}=\phi\topp{i}\si{s}=\phi$ $\forall i$. We stress, however, that the above form of the magnetic dipole operator does not apply to \emph{pseudo}-spins for their effective interactions with external fields; see \sref{sec:wires} for an example involving pseudo-spins.

\subsection{effective spin--spin interactions}
\label{sec:effspinspin}

Inserting the lattice normal-mode expansion~\eref{eq:normalmodes} and~\eref{eq:localcoord} into~\eref{eq:Hint}, we can write the interaction Hamiltonian as
\begin{equation}
	\label{IntHam}
	\mathcal{H}\si{I} =
	\sum_{i=1}^N \sum_{\ell\in\{0,X,Y,Z\}} \Bigg[
		c_{\ell}\topp{i}\cos(\omega\si{I}t+\phi\topp{i}\si{c})
		+\sum_{m=1}^{3 N} 2\hbar\Omega_{im\ell}(\hat{a}_m+\hat{a}\dagg_m)
		\cos(\omega\si{I}t+\phi\topp{i}\si{s})
		\Bigg]\hat{\sigma}_{\ell}\topp{i},
\end{equation}
where we have dropped the approximation symbol and introduced $\Omega_{i m\ell} = \frac{q_{0m}}{2\hbar}\sum_{\mu=1}^3 O_{i\mu m}\vec{m}_i^{\mu}\cdot\vec{s}_{\ell}\topp{i}$.
It is common to transform into the interaction picture to assess the dynamics induced by such an interaction Hamiltonian. In this transformation, the field-free Hamiltonian $\mathcal{H}_0=\sum_{m=1}^{3N} \hbar\omega_m(\hat{a}\dagg_m \hat{a}_m+\frac12) + \hbar\omega_{\uparrow\downarrow} \sum_{i=1}^N\hat{\sigma}_Z\topp{i}$  leads to a time dependence of the operators in~\eref{IntHam}:
\begin{eqnarray}
\fl	\hat{a}_m \mapsto \hat{a}_m e^{-\rmi\omega_m t}
	\qquad
	\hat{a}\dagg_m \mapsto \hat{a}\dagg_m e^{\rmi\omega_m t} &
	\qquad
	\hat{\sigma}_0\topp{i} \mapsto \hat{\sigma}_0\topp{i}
	\qquad
	\hat{\sigma}_Z\topp{i} \mapsto \hat{\sigma}_Z\topp{i}\nonumber\\
\fl	\hat{\sigma}_X\topp{i} \mapsto \hat{\sigma}_X\topp{i}\cos(\omega_{\uparrow\downarrow}t)-\hat{\sigma}_Y\topp{i}\sin(\omega_{\uparrow\downarrow}t) &
	\qquad
	\hat{\sigma}_Y\topp{i} \mapsto \hat{\sigma}_X\topp{i}\sin(\omega_{\uparrow\downarrow}t)+\hat{\sigma}_Y\topp{i}\cos(\omega_{\uparrow\downarrow}t).
\end{eqnarray}
The terms involving $\hat{\sigma}_X\topp{i}$ and $\hat{\sigma}_Y\topp{i}$ can lead either to spin flips without affecting the motion (``carrier''-transitions, mediated by $c_X\topp{i} \hat{\sigma}_X\topp{i}$ and $c_Y\topp{i} \hat{\sigma}_Y\topp{i}$ and resonant at the frequency difference $\omega_{\uparrow \downarrow}$ between the pseudo-spin states) or to interactions that couple spins and motion (``sideband''-transitions or M{\o}lmer--S{\o}rensen interactions~\cite{Moelmer1999a}, mediated by $\vec{s}_X\topp{i} \hat{\sigma}_X\topp{i}$ and $\vec{s}_Y\topp{i} \hat{\sigma}_Y\topp{i}$ and resonant around $\omega_{\uparrow \downarrow} \pm \bar{\omega}_{\mu}$); the latter will dominate if they are not driven too strongly and $|\omega\si{I}-\omega_{\uparrow\downarrow} \pm \bar{\omega}_{\mu}|\ll|\omega\si{I}-\omega_{\uparrow\downarrow}|$ for one of the signs in $\pm$. Here we concentrate instead on a drive with frequency $|\omega\si{I}-\bar{\omega}_3| \ll \bar{\omega}_3 \ll \omega_{\uparrow \downarrow}$, close to one of the three bare eigenfrequencies of the uncoupled ion sites (we have chosen $\mu=3$ without restricting generality). In this case we can neglect the terms in $\vec{s}_X\topp{i}$ and $\vec{s}_Y\topp{i}$ as they are far off-resonant, and  all $c_{\ell}\topp{i}$ by careful design of the experiment (see \sref{sec:wires}). The interaction Hamiltonian in the interaction picture thus reduces to a coherent drive
\begin{equation}
\label{IntRWA}
	\mathcal{H}\si{I}\se{int} \approx
	\hbar \sum_{i=1}^N \sum_{m=1}^{3 N} (\hat{a}_m e^{-\rmi(\delta_m t-\phi\topp{i}\si{s})}+\hat{a}\dagg_m e^{\rmi(\delta_m t-\phi\topp{i}\si{s})})
	(\Omega_{i m 0}\hat{\sigma}_0\topp{i} + \Omega_{i m Z}\hat{\sigma}_Z\topp{i})
\end{equation}
after a second rotating-wave approximation, with the detunings $\delta_m \equiv \omega_m-\omega\si{I}$.
Equation~\eref{IntRWA} can be exactly integrated via a Magnus expansion~\cite{Roos2008,Schneider2011} to yield the unitary evolution operator
\begin{eqnarray}
\label{UniOpe}
\fl	\hat{U}\si{I}\se{int}(t) &=&
	\exp\Bigg[
		\sum_{i=1}^N \sum_{m=1}^{3 N}
		\frac{1-e^{\rmi \delta_m t}}{\delta_m}e^{-\rmi \phi\topp{i}\si{s}} \hat{a}\dagg_m (\Omega_{im0} \hat{\sigma}_0\topp{i}+\Omega_{imZ}\hat{\sigma}_Z\topp{i})
		- \text{h.c.}
	\Bigg] \nonumber \\
\fl	&\times& \exp \Bigg[\rmi \sum_{i,j=1}^N \cos\phi\si{s}^{ij}
		\sum_{m=1}^{3 N}
		(\Omega_{im0} \hat{\sigma}_0\topp{i}+\Omega_{imZ}\hat{\sigma}_Z\topp{i})(\Omega_{jm0} \hat{\sigma}_0\topp{j}+\Omega_{jmZ}\hat{\sigma}_Z\topp{j})
	\frac{\delta_m t-\sin(\delta_m t)}{\delta_m^2}
	\Bigg],
\end{eqnarray}
with $\phi\si{s}^{ij} = \phi\topp{i}\si{s}-\phi\topp{j}\si{s}$.
The first exponent describes a set of time-dependent coherent displacements to all normal modes that can entangle the motion with the internal pseudo-spin states. The second exponent constitutes a phase that depends on pairs of spins and can be interpreted as a \emph{spin--spin interaction}. For a faithful simulation of interacting spins it is desirable that \textit{(A)} the first term should be as close as possible to the identity operator, in order to avoid populating vibrational excitations, and \textit{(B)} the second term should provide sizable phases for desired inter-ion couplings with $i\neq j$, as these represent the spin--spin interactions. It can be shown from the expression above or by the use of a canonical transformation \cite{Porras2004} that \textit{(A)} can be approximately met as long as $|\Omega_{im\ell}| \ll |\delta_m|$ for all $(i,m,\ell)$.

The above restrictions do not limit the time scale for simulations, as long as one assumes that sufficiently strong couplings can be induced by lasers or microwave field gradients. However, the energy scale of nearest-neighbor Coulomb interactions also plays an important role in determining simulation time scales, but this dependence is hidden in the normal-mode coefficients $O_{i \mu m}$ of~\eref{eq:normalmodes}. To illustrate this point, we assume that $\Omega_{im0}=0$ for all normal modes $m$ and sites $i$ [see~\eref{eq:magneticSC} for an example]; but what we show below also holds for  more general cases. Assuming negligible displacements [see point (A) above] the unitary evolution operator thus simplifies to
\begin{eqnarray}
	\label{eq:Usimp}
	\hat{U}\si{I}(t) &=&
	\exp \left[\rmi \sum_{i,j=1}^N
		\hat{\sigma}_Z\topp{i} \hat{\sigma}_Z\topp{j}\cos\phi\si{s}^{ij}
		\sum_{m=1}^{3 N}
		\Omega_{imZ}\Omega_{jmZ}
		\times \frac{\delta_m t-\sin(\delta_m t)}{\delta_m^2}
	\right]\nonumber\\
	&\approx& \exp \left[\rmi t \sum_{i,j=1}^N
		\hat{\sigma}_Z\topp{i} \hat{\sigma}_Z\topp{j}\cos\phi\si{s}^{ij}
		\sum_{m=1}^{3 N}
		\frac{\Omega_{imZ}\Omega_{jmZ}}{\delta_m}
	\right]
	\text{\ for\ } t\gg\sup_m|\delta_m^{-1}|.
\end{eqnarray}
In the ``stiff'' lattice limit (see page~\pageref{stiff}) we choose the drive frequency $\omega\si{I}$ close to one of the bare frequencies, say $\bar{\omega}_3$, such that the detunings $\delta_m$ will be much smaller for normal modes in this set than for the other normal modes; consequently, the sum over modes $m$ in~\eref{eq:Usimp} can be restricted to an ``active'' set of $N$ modes clustered around $\bar{\omega}_3$. If we further choose the drive frequency such that $\bar{\delta}_3=\bar{\omega}_3-\omega\si{I}$ is much larger than the spread of the normal mode frequencies in the active group, the series expansion
\begin{equation}
	\label{eq:phaseseries}
	\frac{1}{\delta_m}
	=\frac{1}{\bar{\delta}_3}
	- \frac{\omega_m^2-\bar{\omega}_3^2}{2\bar{\omega}_3\bar{\delta}_3^2}
	+\mathcal{O}[(\omega_m^2-\bar{\omega}_3^2)^2]
\end{equation}
together with the relations in the ``active'' group of normal modes
\numparts
\begin{eqnarray}
	\label{eq:OOsum1}
	\sum_{m=1}^{N} O_{i3m} O_{j3m} &=& \delta_{ij}\\
	\label{eq:OOsum2}
	\sum_{m=1}^{N} O_{i3m} O_{j3m} (\omega_m^2-\bar{\omega}_3^2) &=& \Cc^{33}_{ij}
\end{eqnarray}
\endnumparts
simplifies the unitary evolution~\eref{eq:Usimp} to
\begin{eqnarray}
	\label{eq:Usimp3}
\fl	\hat{U}\si{I}(t) &\approx&
	\exp \left[
		\frac{\rmi \bar{q}_{03}^2t}{4\hbar^2\bar{\delta}_3}
		\sum_{i=1}^N (\vec{m}_i^3\cdot\vec{s}_Z\topp{i})^2\right]\nonumber\\
\fl	&&\times\exp\left[
		-\frac{\rmi \bar{q}_{03}^2 t}{8 \hbar^2\bar{\omega}_3\bar{\delta}_3^2}
		\sum_{i,j=1}^N
		\Cc^{33}_{i j}\cos\phi\si{s}^{ij}
		(\vec{m}_i^3\cdot\vec{s}_Z\topp{i})(\vec{m}_j^3\cdot\vec{s}_Z\topp{j})
		\hat{\sigma}_Z\topp{i} \hat{\sigma}_Z\topp{j}
	\right],
\end{eqnarray}
where we have further approximated $q_{0 m}\approx \bar{q}_{03}=\sqrt{\hbar/(2M\bar{\omega}_3)}$. The first term in~\eref{eq:Usimp3} is a global phase; it is the second term that mediates an effective spin--spin coupling on the lattice of ions. It can be interpreted as the evolution under an effective spin--spin coupling Hamiltonian due to the driving of local mode $\mu=3$,
\begin{equation}
	\label{eq:spinspinHeff}
	\mathcal{H}_{33}\se{eff} =
		\sum_{i,j=1}^N
		J^{33}_{i j}
		\hat{\sigma}_Z\topp{i} \hat{\sigma}_Z\topp{j}
\end{equation}
where the effective spin--spin coupling coefficients are
\begin{equation}
	\label{eq:spinspinJeff}
	J^{33}_{i j} =
		\frac{	\bar{q}_{03}^2 \Cc^{33}_{i j}\cos\phi\si{s}^{ij}(\vec{m}_i^3\cdot\vec{s}_Z\topp{i})(\vec{m}_j^3\cdot\vec{s}_Z\topp{j})}{8\hbar \bar{\omega}_3\bar{\delta}_3^2}.
\end{equation}
We conclude that to lowest order the strength of the spin--spin coupling between two ions is determined by the geometric overlaps $(\vec{m}_i^3\cdot\vec{s}_Z\topp{i})$ and $(\vec{m}_j^3\cdot\vec{s}_Z\topp{j})$ of the ``active'' local modes of vibration with the direction of the state-dependent force, as well as by the real-space Coulomb coupling strength $\Cc^{33}_{i j}$ between the ions moving along these local modes [see~\eref{eq:ddcoupling} and~\eref{eq:pot2}]. The relative phases $\phi\si{s}^{ij}$ of the driving forces can be used to modulate the coupling strengths. These observations are used in \sref{sec:Kitaev} to construct a quantum simulator on a lattice of trapped ions. Equations~\eref{eq:spinspinHeff} and~\eref{eq:spinspinJeff} faithfully describe the evolution of the system under the following conditions:
\begin{itemize}
	\item The vibrational band structure of the trapped ions must consist of clearly distinct bands which can be addressed individually (see page~\pageref{stiff} for the conditions for ``stiff'' trapping). 
	\item The state-dependent force must be driven at a small detuning $\delta$ from one of these bands: $\delta$ must be large enough such that~\eref{eq:phaseseries} is valid for this band, but small enough such that the contributions to~\eref{eq:spinspinJeff} from other bands are negligible.
	\item The amplitude of the state-dependent force must be small enough such that it does not significantly excite ion vibrations. The condition $|\Omega_{i m\ell}|\ll|\delta_m|$ given above implies that the state-dependent forces must be weaker than the force scale $\hat{F}\sim\hbar|\delta|/q_{0}$ for the addressed band.
\end{itemize}

The above arguments can be made in a very similar fashion for $\hat{\sigma}_X\topp{i}\hat{\sigma}_X\topp{j}$ and $\hat{\sigma}_Y\topp{i}\hat{\sigma}_Y\topp{j}$ interactions by considering the interaction Hamiltonian~\eref{IntHam} with $\omega_I \approx \omega_{\uparrow \downarrow} \pm \bar{\omega}_{\mu}$ in the appropriate basis $\ket{\pm} = (\upket \pm e^{\rmi \chi} \downket)/\sqrt{2}$, where the considered spin--spin interaction is diagonal. The only slight complication can arise from carrier terms proportional to $c_X\topp{i}$ and $c_Y\topp{i}$ that are detuned by roughly the motional eigenfrequencies. For detunings from the sidebands on the order of the dipole--dipole interactions and correspondingly small drive strengths, however, these carrier terms can be safely neglected.

To summarize this section, we have considered general effective spin--spin interactions in the limit of ``stiff'' ion trapping. We have shown that even in a lattice, the spin--spin coupling strength of any two ions depends on the dipolar Coulomb coupling between these two ions. To avoid appreciable entanglement between (pseudo)spins and ion motion, the detunings of driving fields need to be larger than the couplings they induce. This latter finding agrees with other work on simulation with trapped ions~\cite{Porras2006b}. At the same time, the detunings cannot be much larger than the couplings between nearest neighbors, which determine the finer structure of the normal-mode spectrum around the frequencies of the uncoupled (``bare'') motion of an ion tightly bound in one of the trapping wells (see \sref{sec:Kitaev} for a concrete example). These requirements impose stringent bounds on the time scales necessary to perform simulations. For example, in~\cite{Brown11:Coupled-quantiz} two ions  at a distance of $40\,\mu\text{m}$ exhibited an exchange splitting of approximately $3\,\text{kHz}$, barely sufficient to demonstrate a few energy exchanges before ion heating profoundly altered the motion. Simulations that need to progress adiabatically with respect to this exchange period will therefore be experimentally challenging and may require reducing anomalous heating below what was measured in~\cite{Brown11:Coupled-quantiz}.

\section{Kitaev model}
\label{sec:Kitaev}

As an example of how to use the results of sections~\ref{sec:cover-plane-boundary} and~\ref{sec:spinspin} in the design of a quantum simulator, we construct an implementation of the hexagonal Kitaev model~\cite{Kitaev2006} with microtrapped ions. In its ideal form, this exactly solvable two-dimensional spin model has a topologically ordered ground state with anyonic excitations, which makes it extraordinarily interesting for study in a quantum simulator with individual access to the constituent degrees of freedom.

\subsection{model and implementation}

\begin{figure}
	\centering
	\includegraphics[width=8cm]{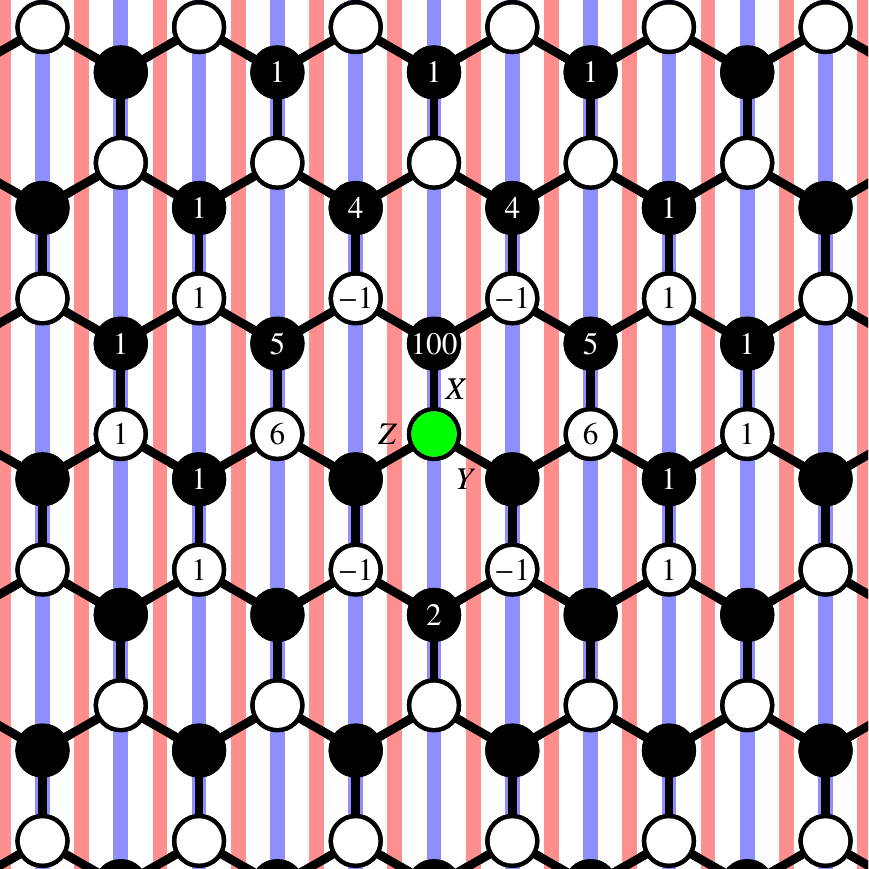}
	\includegraphics[width=12mm]{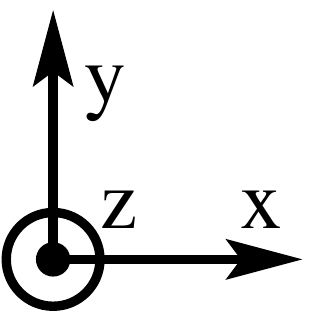}
	\caption{Dipole--dipole interactions~\eref{eq:dipdip1} with the central (green) site due to the vibrational directions of~\eref{eq:vibrations} for the case of $\mu=X$, expressed as percentages of the dominant coupling [equal to $1.99\times Q^2/(4\pi\epsilon_0 d^3)$, see (\ref{eq:Vnn}); values below 1\% of this are not shown]. Figures~\ref{fig:bandstructure} and~\ref{fig:densityofstates} show the vibrational band structure induced by these couplings. The couplings for $Y$ ($Z$) are found by rotating this figure by $120^{\circ}$ ($240^{\circ}$) clockwise. Red and blue wires are described in \sref{sec:wires}.}
	\label{fig:couplings}
\end{figure}

The hexagonal Kitaev model~\cite{Kitaev2006} has the Hamiltonian
\begin{equation}
	\label{eq:Kitaev}
	\mathcal{H}\si{Kitaev} =
	-J_X \sum_{X\text{-links}} \hat{\sigma}_X\topp{i} \hat{\sigma}_X\topp{j}
	-J_Y \sum_{Y\text{-links}} \hat{\sigma}_Y\topp{i} \hat{\sigma}_Y\topp{j}
	-J_Z \sum_{Z\text{-links}} \hat{\sigma}_Z\topp{i} \hat{\sigma}_Z\topp{j}
\end{equation}
defined on a honeycomb lattice of spin-1/2 particles, where the lab-frame bond vectors refer to Fig.~\ref{fig:couplings}:
\begin{equation}
	\label{eq:latticedirections}
	\vec{\Delta}_X = d\{0,1,0\}
	\qquad
	\vec{\Delta}_Y = d\{\sqrt{3},-1,0\}/2
	\qquad
	\vec{\Delta}_Z = d\{-\sqrt{3},-1,0\}/2.
\end{equation}
In this way the Hamiltonian~\eref{eq:Kitaev} associates each real-space bond direction~\eref{eq:latticedirections} with a spin quantization direction; however, it is important to keep in mind that the bond directions and the associated spin quantization directions are not \emph{a priori} related (see \ref{app:coordinates}).
The ions are located on two sublattices $\mathbb{L}_{\latticeA}$ and $\mathbb{L}_{\latticeB}$, as shown in \fref{fig:couplings}. Neighboring ions are a distance $d$ apart.

The form of~\eref{eq:Kitaev} is exactly that of~\eref{eq:spinspinHeff} summed over three concurrent driving force fields. As these driving fields will be at very different frequencies, they can be applied simultaneously in order to drive the full Hamiltonian~\eref{eq:Kitaev}. What is therefore needed in order to implement the Kitaev model is a set of ``bare'' vibrational directions of the ions such that the couplings $\Cc_{ij}^{\mu\nu}$, and therefore the effective spin--spin couplings~\eref{eq:spinspinJeff}, match the particular geometry of the three terms in~\eref{eq:Kitaev}.

\begin{figure}
	\centering
	\includegraphics[width=15cm]{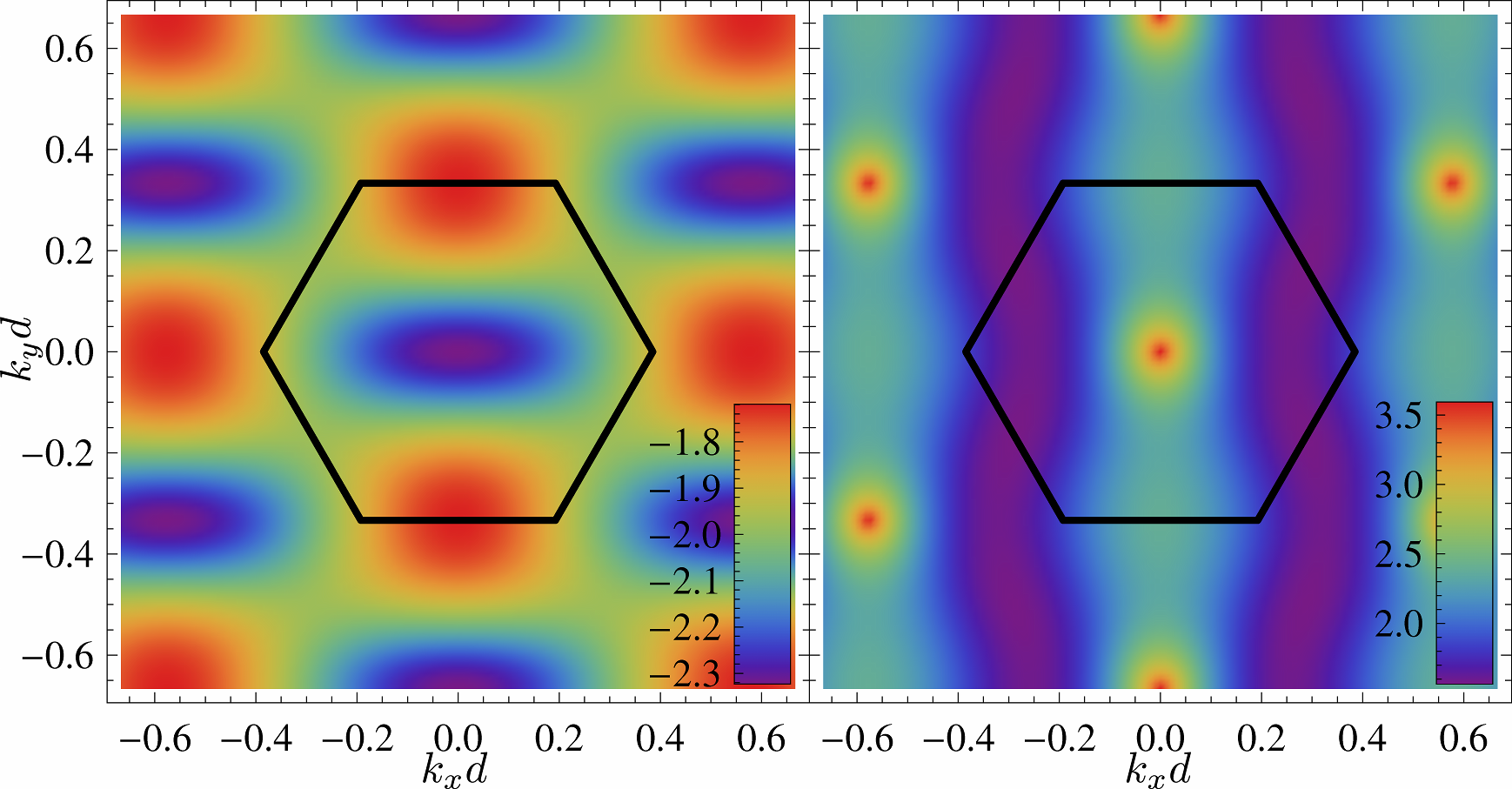}
	\caption{Vibrational band structure due to the dipole--dipole interactions of vibrations along one of the sets of axes in~\eref{eq:vibrations} ($\mu=X$, see \fref{fig:couplings}), corresponding to the density of states shown in \fref{fig:densityofstates}. In the lower band (left) neighboring ions move out of phase; in the upper band (right) they move in phase. The first Brillouin zone is drawn in black. Frequencies (colors) are given in units of $\omega_0$ (see \fref{fig:densityofstates}).}
	\label{fig:bandstructure}
\end{figure}

\begin{figure}
	\centering
	\includegraphics[height=5cm]{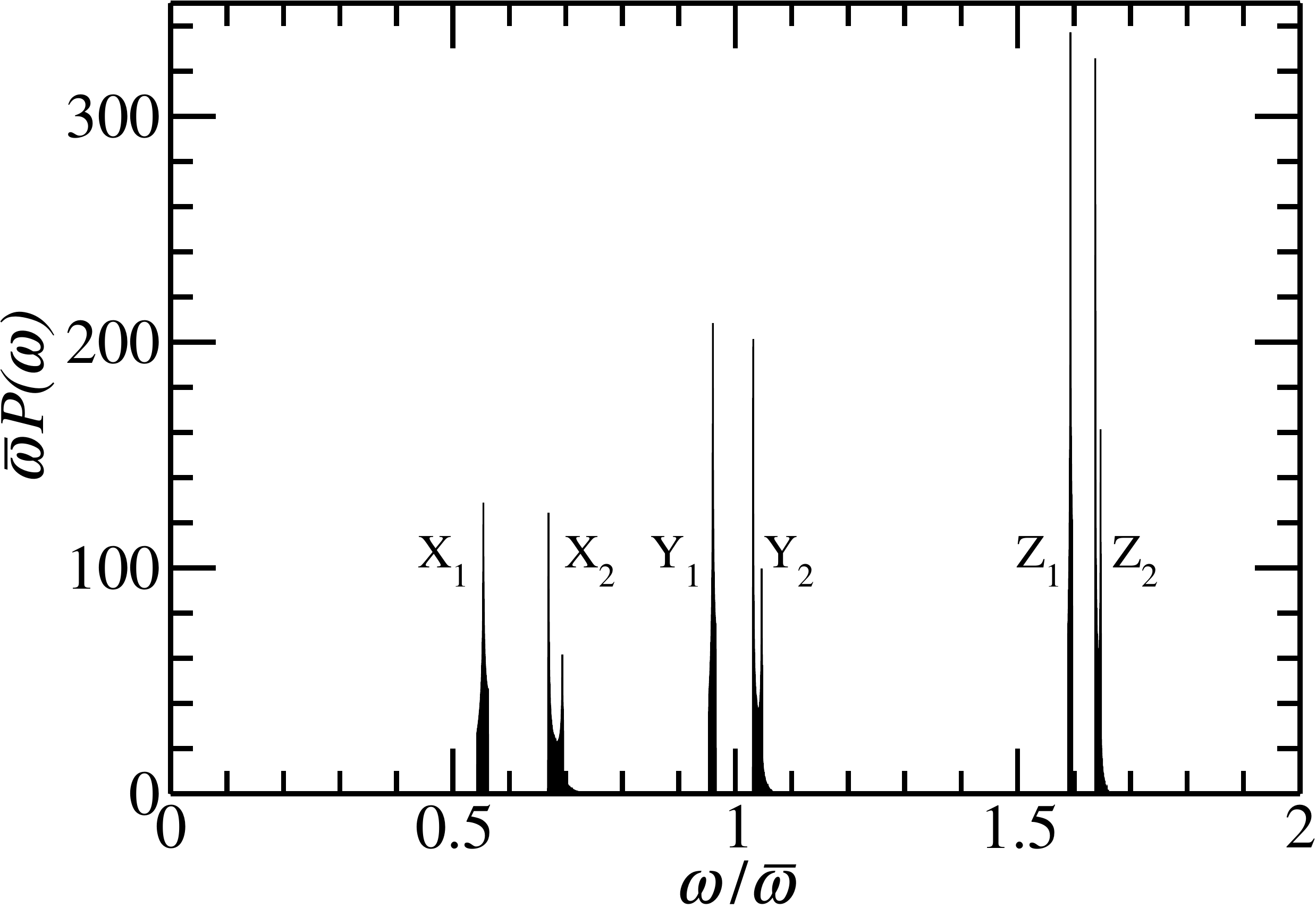}
	\includegraphics[height=5cm]{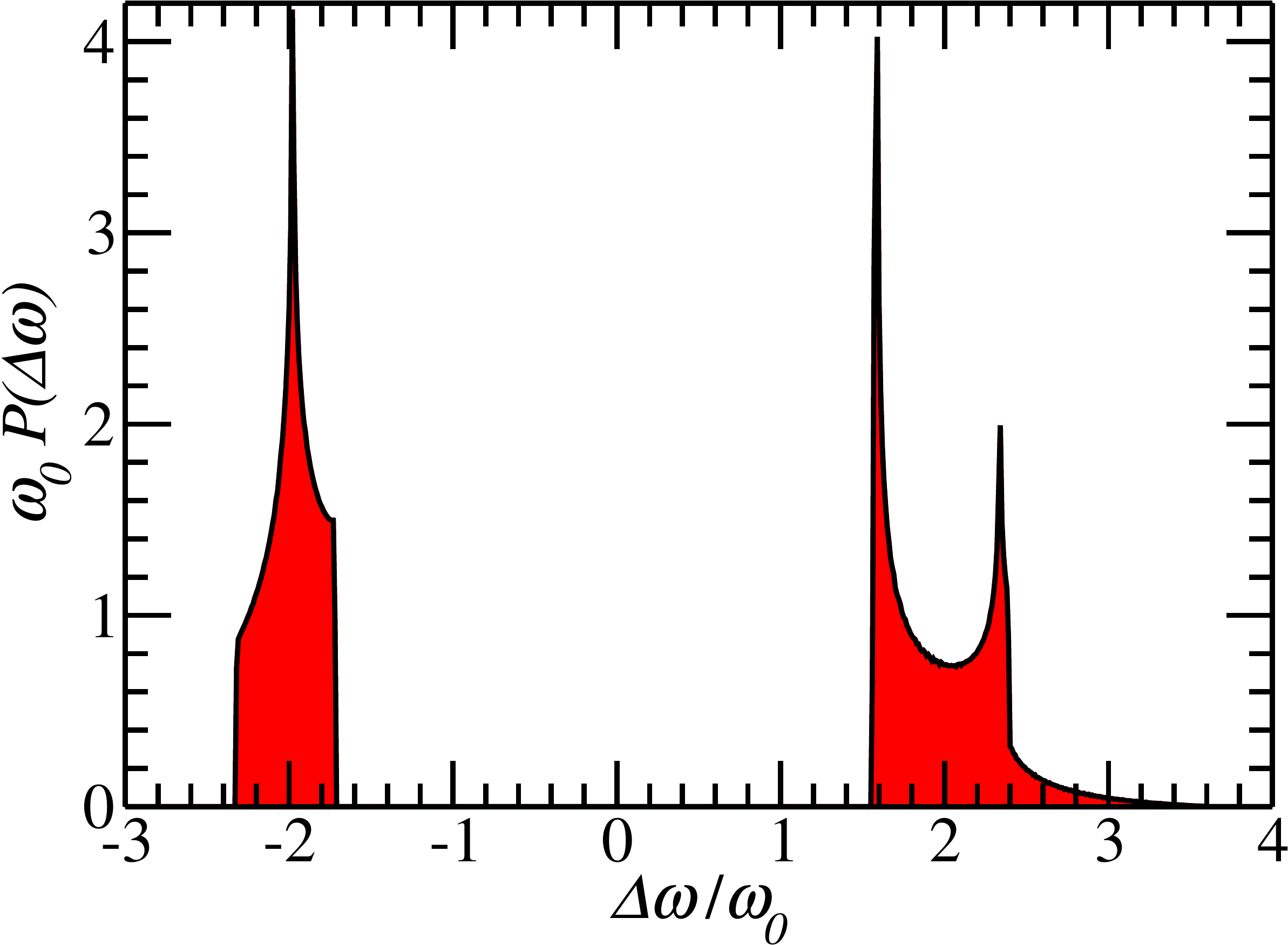}
	\caption{Density of states of the vibrational bands (\fref{fig:bandstructure}) due to the dipole--dipole interactions of vibrations along the sets of axes in~\eref{eq:vibrations}, shown in \fref{fig:couplings}. Left: the six bands consisting of three off-resonant doublets ($\mu=X,Y,Z$) with center frequencies (bare trap eigen-frequencies) split by the golden ratio (see \sref{sec:KitaevElectrode}); $\bar{\omega}=(\bar{\omega}_X\bar{\omega}_Y\bar{\omega}_Z)^{1/3}$ and $\omega_{0Y}/\bar{\omega}=0.02$ (much larger than in a realistic experiment). Right: zoom of one of the doublets. The two bands detailed in \fref{fig:bandstructure} are clearly visible, separated by $\approx4\omega_0$, and show the extent to which the couplings of \fref{fig:couplings} are dominated by the desired nearest-neighbor couplings. The scale of the bands is $\omega_{0\mu}=Q^2/(8\pi\epsilon_0\bar{\omega}_{\mu} M d^3)$.}
	\label{fig:densityofstates}
\end{figure}

We choose the ion trapping height to be half of the inter-ion distance, $h=d/2$, and the orthonormal principal axes of vibration for ions on the two sublattices as
\begin{eqnarray}
	\label{eq:vibrations}
	\vec{m}_{\latticeA}^X = \{0,2,\sqrt{2}\}/\sqrt{6} & \vec{m}_{\latticeB}^X = \{0,-2,\sqrt{2}\}/\sqrt{6}\nonumber\\
	\vec{m}_{\latticeA}^Y = \{\sqrt{3},-1,\sqrt{2}\}/\sqrt{6} & \vec{m}_{\latticeB}^Y = \{-\sqrt{3},1,\sqrt{2}\}/\sqrt{6} \nonumber\\
	\vec{m}_{\latticeA}^Z = \{-\sqrt{3},-1,\sqrt{2}\}/\sqrt{6} \qquad\qquad & \vec{m}_{\latticeB}^Z = \{\sqrt{3},1,\sqrt{2}\}/\sqrt{6}.
\end{eqnarray}
This particular choice of axes of vibration has the property that dipole--dipole couplings of the sort of $\Cc_{ij}^{\mu\mu}$ (\emph{i.e.}, coupling the $\vec{m}_i^{\mu}$ vibration of the ion at $\vec{R}_{0i}$ with the $\vec{m}_j^{\mu}$ vibration of the ion at $\vec{R}_{0j}$) are strongly dominated by the nearest-neighbor couplings required by the Kitaev model~\eref{eq:Kitaev}, shown in \fref{fig:couplings} for $\mu=X$. These couplings can be calculated from~\eref{eq:dipdip1} (in the absence of a cover plane); the resulting nearest-neighbor terms in the dipole--dipole coupling part of the Coulomb potential~\eref{eq:pot2} are
\begin{equation}
	\label{eq:Vnn}
	\fl
	V\si{nn}
	= \sum_{\mu,\nu\in\{X,Y,Z\}} \frac{Q^2}{4\pi\epsilon_0 d^3} \left[
	\frac{52-3\sqrt{2}}{24}\delta_{\mu,\nu}
	+\frac{3\sqrt{2}-4}{48}(1-\delta_{\mu,\nu})
	\right]
	\sum_{i\in\mathbb{L}_{\latticeA}}
	\vibc_i^{\mu}\vibc_{i+\vec{\Delta}_{\nu}}^{\mu}.
\end{equation}
In addition, there are dipole--dipole couplings to neighbors that are further away and that turn out to be larger than the off-diagonal terms ($\mu\neq\nu$) in~\eref{eq:Vnn}. The vibrational normal-mode band structure due to all of these dipole--dipole couplings is shown in \fref{fig:bandstructure}, with the effective density of states shown in \fref{fig:densityofstates}. It consists of two bands, in which neighboring ions oscillate in-phase (upper band) and out-of-phase (lower band), and whose small frequency spread is indicative of the dominance of the nearest-neighbor coupling over all other couplings.

Many dipole--dipole couplings of the sort of $\vibc_i^{\mu}\vibc_j^{\nu}$ with $\mu\neq\nu$ are nonzero in this configuration; however they do not lead to effective spin--spin couplings if the underlying trap frequencies along the directions $\vec{m}_i^{\mu}$ and $\vec{m}_j^{\nu}$ are strongly off-resonant (see sections~\ref{sec:effspinspin} and~\ref{sec:KitaevElectrode}). Thus neglecting any $\mu\neq\nu$ couplings, the effective spin--spin Hamiltonian that is constructed from $\mu=X$ Coulomb interactions is approximately
\begin{eqnarray}
	\label{eq:KitaevApproxZ}
	-\mathcal{H}_X/J_X =& \sum_{i\in\mathbb{L}_{\latticeA}}
	\hat{\sigma}_X\topp{i} \left[
		\hat{\sigma}_X\topp{i+\vec{\Delta}_X}
		+0.05 \left( \hat{\sigma}_X\topp{i-2\vec{\Delta}_Y} + \hat{\sigma}_X\topp{i-2\vec{\Delta}_Z} \right)
		+\ldots\right]\nonumber\\
	&+\frac{1}{2}\sum_{i\in\mathbb{L}_{\latticeA}\cup\mathbb{L}_{\latticeB}}
	\hat{\sigma}_X\topp{i} \left[ 0.06 \left( \hat{\sigma}_X\topp{i+\vec{\Delta}_Y-\vec{\Delta}_Z} + \hat{\sigma}_X\topp{i+\vec{\Delta}_Z-\vec{\Delta}_Y} \right)
	+\ldots\right],
\end{eqnarray}
where the first sum contains couplings between the different lattices while the second sum contains couplings within the lattices; numerical prefactors for the perturbing terms are used for brevity, as in \fref{fig:couplings}. Here we have assumed for simplicity that all ions are simultaneously pushed by the same state-dependent force with equal phase. The Hamiltonians $\mathcal{H}_Y$ and $\mathcal{H}_Z$ are found from~\eref{eq:KitaevApprox} through rotations by $\pm120^{\circ}$, and the total effective spin Hamiltonian is
\begin{equation}
	\label{eq:KitaevApprox}
	\mathcal{H}'=-J_X\mathcal{H}_X-J_Y\mathcal{H}_Y-J_Z\mathcal{H}_Z,
\end{equation}
where $J_X$, $J_Y$, and $J_Z$ are effective coupling constants containing the diagonal coupling strength, the physical prefactors, as well as the mechanisms used for achieving these effective spin--spin couplings (see \sref{sec:wires}). The topic of whether or not this slightly perturbed Hamiltonian~\eref{eq:KitaevApprox} exhibits the same interesting topological phases as the ideal Hamiltonian~\eref{eq:Kitaev}, at zero or finite~\cite{Nussinov2008,Hastings2011} temperature, is beyond the scope of this article. We mention, however, that if the perturbative terms of~\eref{eq:KitaevApprox} will be deemed too strong, they can be reduced further by driving the different wires with different relative phases or amplitudes [see~\eref{eq:spinspinJeff}].

The presented configuration of trapping height and vibrational axes nearly maximizes the desired dipole--dipole couplings at the same time as it nearly mimimizes all undesired couplings. By numerical optimization we can identify a configuration that performs a few percent better than~\eref{eq:vibrations}, but we have not been able to obtain an analytical description of this configuration.

\subsection{surface-electrode trap design}
\label{sec:KitaevElectrode}

\begin{figure}
	\centering
	\includegraphics[width=7.5cm]{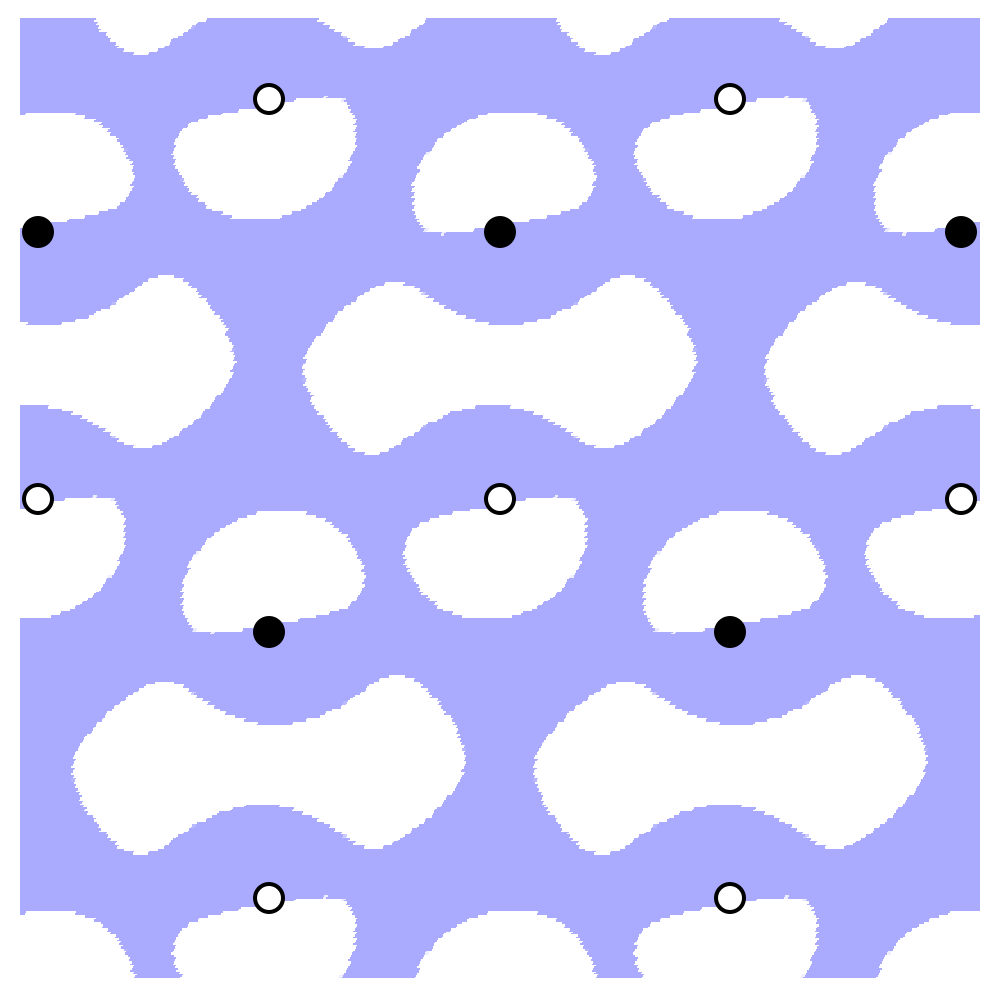}
	\includegraphics[width=7.5cm]{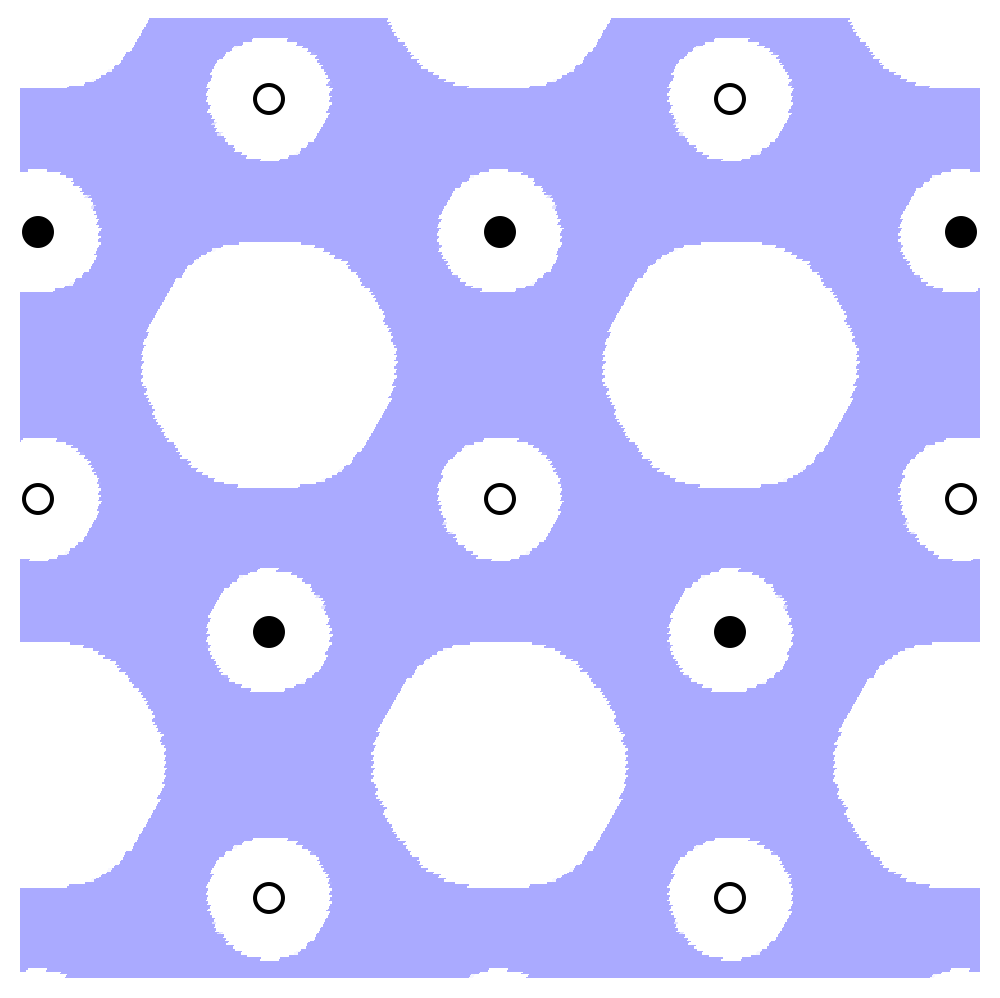}
	\caption{Left: optimized rf (blue) and dc (white) electrodes for the constraints of \sref{sec:KitaevElectrode}, with no spurious traps. Dimensionless trap curvatures are $\kappa=0.080$. Right: optimized electrodes for a honeycomb lattice with out-of-plane quadrupole confinement, trapping height $h=d/2$, and cover plane at $H=50d$. Dimensionless trap curvatures are $\kappa=0.102$. Coordinates as in \fref{fig:couplings}.}
	\label{fig:electrodes}
\end{figure}

To have maximally incommensurate vibrational frequencies along the normal mode axes~\eref{eq:vibrations} we choose them in the golden ratio $\omega_X:\omega_Y:\omega_Z=\phi^{-1}:1:\phi$ with $\phi=(1+\sqrt{5})/2$. We use the algorithm of Ref.~\cite{Schmied2009} to find an rf surface-electrode pattern that will generate an infinite honeycomb lattice of exactly such microtraps, with the following constraints:
\begin{itemize}
	\item The unit cell of the electrode pattern is defined by the vectors $\vec{a}=d\{\sqrt{3},0,0\}$ and $\vec{b}=d\{\sqrt{3}/2,3/2,0\}$.
	\item The ion positions within the unit cell define the sublattices $\mathbb{L}_{\latticeA}$ and $\mathbb{L}_{\latticeB}$: $\vec{R}_{0\latticeA}=d\{0,0,1/2\}$ and $\vec{R}_{0\latticeB}=d\{\sqrt{3},1,1/2\}$.
	\item The gradient of the rf electric potential generated by the surface electrodes must vanish at the ion positions in order to have minima of the rf pseudo-potential.
	\item The principal axes of the second derivative tensors of the rf electric potential at the ion positions are aligned with the directions given in~\eref{eq:vibrations}, with eigenvalues proportional to $\{\phi^{-1},1,-\phi\}$ in the $\vec{m}_{\latticeA,\latticeB}^X$, $\vec{m}_{\latticeA,\latticeB}^Y$, and $\vec{m}_{\latticeA,\latticeB}^Z$ directions, respectively.
	\item A cover plane is located at a height $H=50d$.
\end{itemize}
The resulting electrode pattern is shown in the left panel of \fref{fig:electrodes}. It generates microtraps at the desired positions with dimensionless curvatures~\cite{Schmied2009} $\kappa=0.080$ and no spurious additional microtraps. This is to be compared with a simple out-of-plane quadrupole honeycomb lattice geometry ($\kappa=0.102$) as in Ref.~\cite{Schmied2009} (see \fref{fig:electrodes}, right panel), which can potentially be deformed during the experiment via dc electrode potentials into satisfying the above constraints. Such dc electrodes might be necessary in any experimental implementation in order to null micro-motion of the ions~\cite{Berkeland1998} induced by manufacturing inaccuracies and stray charges.

\subsection{trap depth and trap loading}

\begin{figure}
	\centering
	\includegraphics[height=5cm]{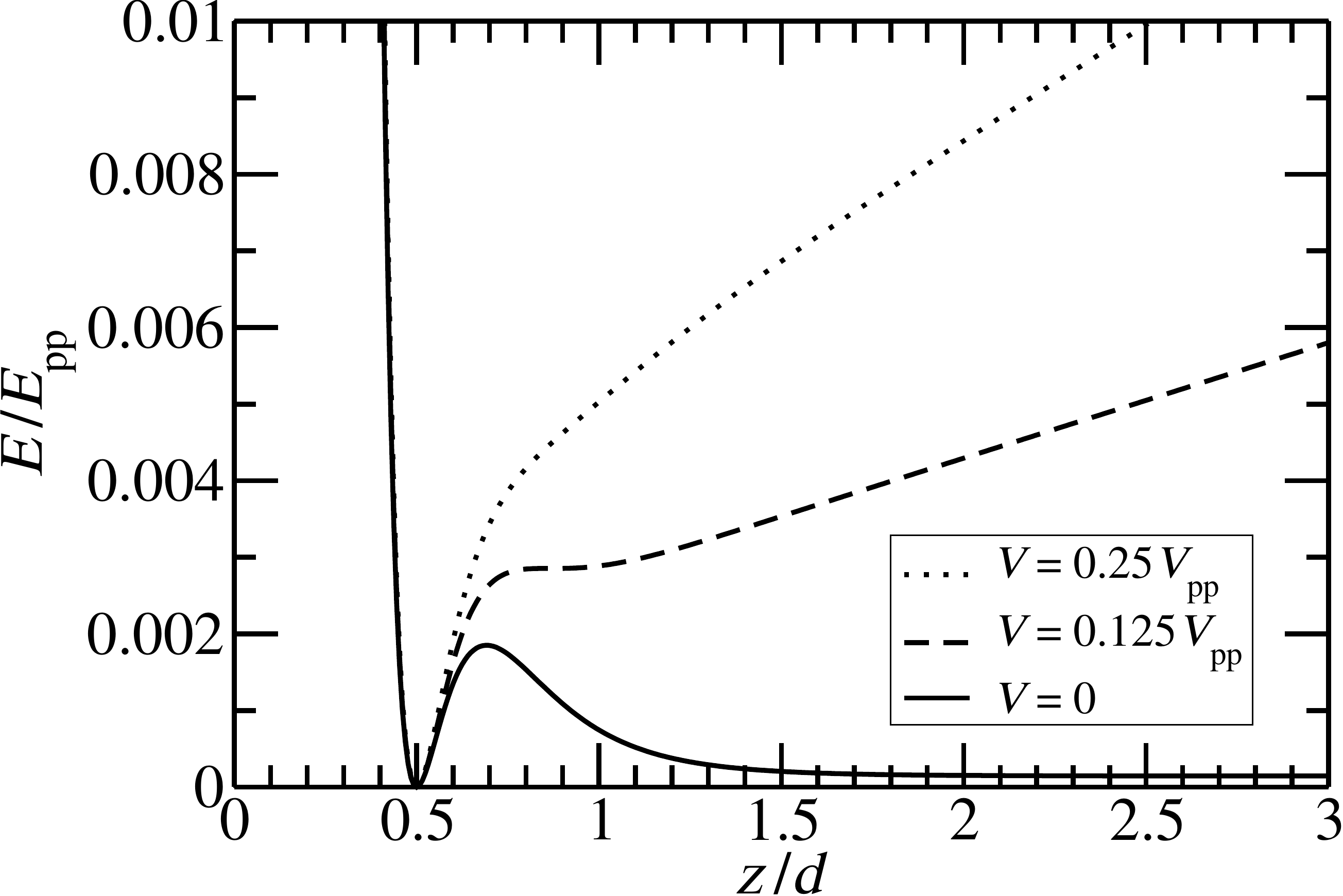}
	\caption{Total potential on a vertical axis through any microtrap formed by the electrode of \fref{fig:electrodes} (right panel). The ponderomotive pseudo-potential is drawn with a solid line, in units of $E\si{pp}=Q^2U\si{rf}^2/(4M\Omega\si{rf}^2d^2)$. Two levels of dc biasing (either through the rf electrode or applying the same bias potential to the dc electrodes and the cover plane; vertically offset to leave the trapping minimum unchanged) are shown, in units of $V\si{pp}=E\si{pp}/Q$. For a bias voltage $V\gtrsim0.125V\si{pp}$ the hexagonal lattice microtraps are the only attractors (trapping zones) for cooled ions.}
	\label{fig:verticalpot}
\end{figure}

The depth of the microtrap lattices generated by the electrodes shown in \fref{fig:electrodes} are rather shallow. For the honeycomb lattice, which is more easily analyzed due to its p6m symmetry, \fref{fig:verticalpot} (solid line) shows the ponderomotive pseudo-potential along a vertical axis through any microtrap, in units of $E\si{pp}=Q^2U\si{rf}^2/(4M\Omega\si{rf}^2d^2)$. For $^9$Be$^+$ ions ($Q=+e$, $M=9\,\text{u}$) trapped with $U\si{rf}=50$\,V and $\Omega\si{rf}=2\pi\times200$\,MHz in a lattice of $d=30\,\mu\mathrm{m}$ ($h=d/2=15\,\mu\mathrm{m}$ and $H=50d=1.5\,\text{mm}$), we have $E\si{pp}=4.7\,\mathrm{eV}=5.5\times10^4k\si{B}$\,K and thus a trap depth (ion-loss barrier) of $0.00185E\si{pp}=8.7\,\mathrm{meV}=101k\si{B}$\,K. This small adiabatic trap depth is likely further reduced by the breakdown of the pseudo-potential approximation near the trap barrier. In order to reliably load these microtraps we can make use of the cover plane (see \sref{sec:screen-effects-cover-plane}): applying a positive dc bias potential to the dc electrodes and the same potential to the cover plane adds an electrostatic potential that pushes the ions towards the rf electrode and increases the trapping well depth (see \fref{fig:verticalpot}). Since this dc potential is equivalent to applying a negative dc bias potential to the rf electrode, its dc electric field at the ion trap sites vanishes (by construction of the rf electrode shape) and it thus does not induce micro-motion~\cite{Berkeland1998}. We find that applying a small dc bias potential of at least $0.125E\si{pp}/Q\approx0.6$\,V is sufficient to make the desired lattice of microtraps the only minima of the total potential (dashed line in \fref{fig:verticalpot}). By applying a stronger bias voltage, the resulting total potential (dotted line in \fref{fig:verticalpot}) is deep enough to trap ions produced by photoionization directly from a hot atomic beam. This bias will simultaneously cause the traps to be shallower in the $x y$ plane.

\subsection{wires for magnetic interaction}
\label{sec:wires}

As described in \sref{sec:spinspin}, effective spin--spin interactions between ions require internal-state-dependent forces to be applied to the ions. In the present model we propose to embed parallel wires below the electrode plane, which generate local magnetic field gradients at the positions of the ions [see~\eref{eq:magneticSC}]. A relatively simple periodic grid of two different types of wires, indicated in red and blue in \fref{fig:couplings}, suffices to implement the spin--spin interactions along all three bond types of the Kitaev model. As explained in \sref{sec:spinspin}, one can induce pairwise $\hat{\sigma}_X^{(\latticeA)} \hat{\sigma}_X^{(\latticeB)}$ and $\hat{\sigma}_Y^{(\latticeA)}\hat{\sigma}_Y^{(\latticeB)}$ interactions by currents at frequencies that are near-resonant to $\omega_{\uparrow \downarrow}\pm\omega_X$ and $\omega_{\uparrow \downarrow}\pm\omega_Y$ (M\o lmer--S\o rensen type interactions~\cite{Soerensen1999}); the $\hat{\sigma}_Z^{(\latticeA)}\hat{\sigma}_Z^{(\latticeB)}$ interactions can be driven with currents that are near-resonant to $\omega_Z$ (phase-gate type interactions~\cite{Leibfried2003b}). Because the three band-manifolds are well separated in frequency (\fref{fig:densityofstates}), the dynamics of the three bond types can be driven simultaneously by currents at separate frequencies that are mutually off-resonant.

The geometry of the wires is determined by the condition that we need to suppress the magnetic field at the position of all ions, in order to have negligible carrier interactions $c_{\ell}\topp{i}$ in~\eref{eq:magneticSC}, while maintaining useful field gradients that couple to their target vibrational directions~\eref{eq:vibrations} to drive spin--spin interactions on all three bonds simultaneously~\cite{Ospelkaus2008,Ospelkaus2011}. The magnetic field of two infinite sets of infinitely long wires parallel to the $y$ axis as in \fref{fig:couplings}, with distance $d\si{w}=d\sqrt{3}/2$ between wires of equal color, is
\begin{equation}
	\label{eq:MagFieldWW}
\fl	\vec{B}\si{w}(x,z) =\frac{\mu_0 I\si{blue}}{d\sqrt{3}} \times \frac{\vec{\hat{x}}\sinh\frac{4\pi z}{d\sqrt{3}}-\vec{\hat{z}}\sin\frac{4\pi x}{d\sqrt{3}}}{\cos\frac{4\pi x}{d\sqrt{3}}-\cosh\frac{4\pi z}{d\sqrt{3}}}
	- \frac{\mu_0 I\si{red}}{d\sqrt{3}} \times \frac{\vec{\hat{x}}\sinh\frac{4\pi z}{d\sqrt{3}}+\vec{\hat{z}}\sin\frac{4\pi x}{d\sqrt{3}}}{\cos\frac{4\pi x}{d\sqrt{3}}+\cosh\frac{4\pi z}{d\sqrt{3}}},
\end{equation}
where $\mu_0$ is the magnetic constant.
The field over the blue wires (\emph{i.e.}, at the ion positions) vanishes at height $h\si{w}$ if the ratio of currents is
\begin{equation}
    \label{eq:CurrentRatio}
    \frac{I_{\rm blue}}{I_{\rm red}}=-\tanh^2 \frac{2\pi  h\si{w}}{d\sqrt{3}}.
\end{equation}
With this current ratio the magnetic-field gradient at the ion positions is
\begin{equation}
	\vec{\nabla}\vec{B}\si{w}(x=n d\si{w},z=h\si{w}) = \frac{4\pi \mu_0 I\si{blue}}{3d^2}\sinh^{-2} \frac{2\pi  h\si{w}}{d\sqrt{3}}
	\left(\begin{array}{ccc}
		0 & 0 & 1 \\
		0 & 0 & 0 \\
		1 & 0 & 0 \\
	\end{array}\right)
\end{equation}
for any $n\in\mathbb{Z}$.
As this magnetic-field gradient decreases rapidly with increasing distance $h\si{w}$ to the ions, one should place the wires as close as possible to the ions. On the other hand, they should not interfere with the trap electrodes. As a reasonable compromise for the following estimates, we can assume that the wires are below the electrodes such that $h\si{w} = d\si{w} =h\sqrt{3}$. The actual $h\si{w}$ in an experiment will probably be dictated by constraints in the micro-fabrication.

We choose the quantization axis of the pseudo-spins of the ions to coincide with its associated bond direction, $\vec{\hat{Z}}=\vec{\Delta}_Z/d$; however, any other choice of $\vec{\hat{Z}}$ will be equally valid, and the experimenter's choice may depend on the available quantization fields. For our choice, $\vec{s}_Z =  \frac{\sqrt{3}}{2}\vec{\hat{z}} \times g \mu\si{B} \frac{4\pi \mu_0 I\si{blue}^{(\bar{\omega}_Z)}}{3d^2}\sinh^{-2} \frac{2\pi  h\si{w}}{d\sqrt{3}}$ for all ions on both sublattices, and with $\Cc_{\latticeA\latticeB}^{Z Z}=\frac{Q^2}{4\pi\epsilon_0 M d^3}\frac{52-3\sqrt{2}}{24}$ [the diagonal term of~\eref{eq:Vnn}] the interaction strength~\eref{eq:spinspinJeff} becomes
\begin{equation}
    \label{eq:Zinteraction}
\fl    J_Z= \frac{\Cc_{\latticeA\latticeB}^{Z Z}(\vec{m}_{\latticeA}^Z \cdot \vec{s}_Z)(\vec{m}_{\latticeB}^Z \cdot \vec{s}_Z)}{16M \bar{\omega}_Z^2\bar{\delta}_Z^2}
	= \frac{\pi^2(52-3\sqrt{2})}{432} \frac{[\bar{q}_{0Z} g \mu\si{B} \mu_0 Q I\si{blue}^{(\bar{\omega}_Z)}]^2}{4\pi\epsilon_0 M \hbar \bar{\omega}_Z \bar{\delta}_Z^2 d^7} \sinh^{-4} \frac{2\pi  h\si{w}}{d\sqrt{3}},
\end{equation}
where $g$ is the effective $g$-factor such that the energy difference between the $\upket$ and $\downket$ pseudo-spin states in a weak constant magnetic field along the quantization axis is $\Delta E_{\uparrow\downarrow} = \hbar\omega_{\uparrow\downarrow}-g \mu\si{B}B_Z$. $I\si{blue}^{(\bar{\omega}_Z)}$ is the current amplitude in the blue wires at frequency $\bar{\omega}_Z+\bar{\delta}_Z$ with $|\bar{\delta}_Z|\ll\bar{\omega}_Z$. With $d=30\,\mu\text{m}$, $g=1$, $M=9\,\text{u}$, $Q=+e$, and $\bar{\omega}_Z=2\pi\times5\,\text{MHz}$, this coupling strength is $J_Z=7.6\,\text{kHz}\times[I\si{blue}^{(\bar{\omega}_Z)}/\text{A}]^2[\bar{\delta}_Z/(2\pi\,\text{kHz})]^{-2}$. To avoid sizable entanglement of the (pseudo)spins with the ion motion, we need to fulfill
\begin{equation}
	\label{eq:noentangle}
	\left|\frac{J_Z}{\hbar \bar{\delta}_Z}\right| \approx 7.6\times
	\frac{[I\si{blue}^{(\bar{\omega}_Z)}/\text{A}]^2}{[\bar{\delta}_Z/(2\pi\,\text{kHz})]^3}
	<1.
\end{equation}
The hexagonal Kitaev model features interesting gapped phases with anyonic excitations of the ground state for example if $|J_x|=|J_y|<|J_z|/2$, in which case the coupling constant of the resulting effective Hamiltonian is $J\si{eff}=J_x^2 J_y^2/(16|J_z|^3)<|J_z|/256$ \{see~\cite{Kitaev2006} for discussions of these phases and their emergence from~\eref{eq:Kitaev}\}. 

While the geometric prefactor of~\eref{eq:Zinteraction} depends on the details of the model, its functional dependences are expected to remain the same for a broad class of wire-driven coupled pseudo-spin models, in particular also for $J_X$ and $J_Y$ of the same system. The exact form of the interactions along the $X$ and $Y$ bonds depends on the transition dipole matrix elements $\vec{\mu}\si{d}=\me{\uparrow\!\!}{\vec{\hat{\mu}}}{\!\!\downarrow}$ which can have components along all spatial directions.\footnote{In the case of a real spin, $\vec{\mu}\si{d}=\frac12 g\mu\si{B}(\vec{\hat{X}}-\rmi\vec{\hat{Y}})$ (see the example on page~\pageref{dipoleop}); but what follows also applies to more general pseudo-spin cases where we can have $\vec{\mu}\si{d}\cdot\vec{\hat{Z}}\neq0$, for example if the pseudo-spin states are hyperfine states of an ion.} The component along the quantization axis $\vec{\hat{Z}}$ is relevant for $\pi$ transitions (where the $\vec{\hat{Z}}$ component of the total angular momentum $\vec{F}$ of the ion does not change, $\Delta m_F=0$) while perpendicular components can be used for $\sigma^\pm$ transitions (during which the $\vec{\hat{Z}}$ component of $\vec{F}$ changes by $\Delta m_F=\pm 1$).
The coupling strengths for $\pi$ transitions are found by scaling~\eref{eq:Zinteraction} to be
\begin{eqnarray}
	\label{eq:XYinteraction}
	J_{X/Y} &=& J_Z
		\left| \frac{\bar{q}_{0 X/Y} (\vec{\mu}\si{d}\cdot\vec{\hat{Z}})}{\bar{q}_{0 Z} g \mu\si{B}} \right|^2
		\frac{I\si{blue}^{(\omega_{\uparrow \downarrow}+\bar{\omega}_{X/Y})} I\si{blue}^{(\omega_{\uparrow \downarrow}-\bar{\omega}_{X/Y})}}{[I\si{blue}^{(\omega_Z)}]^2},
\end{eqnarray}
with currents of amplitude $I\si{blue}^{(\omega_{\uparrow \downarrow}\pm\bar{\omega}_{X/Y})}$ at frequencies $\omega_{\uparrow \downarrow}\pm(\bar{\omega}_{X/Y}+\bar{\delta}_{X/Y})$ that are required to drive M\o lmer--S\o rensen-type interactions~\cite{Soerensen1999}. For $\sigma^{\pm}$-transitions analogous relations hold involving the projections of $\vec{\mu}\si{d}$ along $(\vec{\hat{X}}\pm\rmi\vec{\hat{Y}})/\sqrt{2}$. We conclude that all interactions are similar in magnitude and that the current amplitudes can be used to tune the effective coupling strengths $J_X$, $J_Y$, and $J_Z$ of the Kitaev model simulator~\eref{eq:KitaevApprox}. To drive all bonds simultaneously, a total of five alternating currents at different frequencies are necessary; using the largest allowed value of 1 in~\eref{eq:noentangle}, the maximum rms current each of the blue wires [and red wires, see~\eref{eq:CurrentRatio}] has to sustain is $\sqrt{\langle I\si{blue}^2(t)\rangle} \approx \sqrt{5/2}\times 0.36\,\text{A} \times [\bar{\delta}/(2\pi\,\text{kHz})]^{3/2}$. We recall, however, that~\eref{eq:Zinteraction} and~\eref{eq:XYinteraction} depend very strongly on the vertical distance $h\si{w}$, and even a small reduction in $h\si{w}$ can substantially reduce the required currents.

Expression \eref{eq:Zinteraction} seems to suggest that  for quantum simulators built with the principles described here, decreasing the physical size of the ion-trap lattice (as given by the length scale $d$) will strongly increase the simulation speed, which is given by the effective dynamics of the particular quantum simulator but ultimately proportional to the spin--spin coupling strengths. However, a careful analysis of assumptions and constraints, carried out below for interactions along $Z$ but equally valid for all other effective interactions, disproves this observation. Firstly, if we assume that avoiding ion motion is a significant experimental constraint, a constant ratio $J_Z/(\hbar\bar{\delta}_Z)$ in~\eref{eq:noentangle} implies that the effective spin--spin coupling strength scales as
\begin{equation}
	\label{eq:basicscaling}
	J_Z \propto \frac{[I\si{blue}^{(\bar{\omega}_Z)}]^{2/3}}{\bar{\omega}_Z^{2/3} d^{7/3}}
\end{equation}
for a given ion species and given electrode shapes.
Secondly, assuming the currents to be limited by heat dissipation, an upper bound on $I\si{blue}^{(\bar{\omega}_Z)}$ must scale as $d^{3/2}$.
And lastly, lower bounds for the trap frequency can be found in two ways: (i) to meet our assumption of stiff trapping, \emph{i.e.}, $\omega_{0Z} \ll \bar{\omega}_Z$, we require $\bar{\omega}_Z \gg d^{-3/2}\times|Q|/\sqrt{8\pi \epsilon_0 M}$; and (ii) to use the expansion~\eref{eq:phaseseries} we require $\omega_{0Z}\ll|\bar{\delta}_Z|$, which, combined with the scaling of~\eref{eq:basicscaling} for $\bar{\delta}_Z$ and with the above current scaling, implies a scaling of $d^{-5}$ for the lower bound of $\bar{\omega}_Z$. Together, these bounds imply that the maximal achievable coupling strength~\eref{eq:basicscaling} scales only as $d^{-1/3}$ or even $d^2$, depending on which of the frequency bounds is more stringent.
Since current experimental setups are far from reaching these lowers bounds on $\bar{\omega}_Z$, miniaturization is expected for now to increase the coupling strength faster than these estimates; but the optimal dimension, where the ratio of simulation speed and heating rate (anomalous heating, scaling as $d^{-4}$~\cite{Turchette2000,Deslauriers2006}) is maximized, remains an open question.

\section{Conclusions}

We have discussed modifications to Coulomb potentials and interactions of trapped ions due to the presence of trap electrodes and cover planes. For plane geometries we have treated these modifications rigorously, using the method of image charges. We have found considerable deviations of the long-range behavior from that in free space when the relevant distances are of the order of ion-to-surface distances or larger. Moreover, we have developed a general approach to treating the effective spin--spin interactions of ions trapped in a multi-trap array in the stiff-trapping limit, where dipole--dipole interactions between nearest neighbors produce only small corrections to the bare normal modes of a given trap well. We have shown that effective coupling strengths, and therefore simulation timescales, are determined by the nearest-neighbor dipole--dipole couplings. As an illustration of the versatility and power of this stiff-trap-array approach, we have discussed  a quantum simulation of the hexagonal Kitaev model. We have also addressed several practical challenges, including how the trap depth of the array may be improved so ions created from a thermal source with large kinetic energies can be trapped.

\ack
We thank Miguel Aguado and Ignacio Cirac for discussions on the properties of the Kitaev model, Dave Wineland for discussions and comments on the manuscript, Jonathan Home for discussions on ion traps, and Tobias Sch\"atz and Christian Schneider for discussions on couplings in large arrays.
R.S.\ gratefully acknowledges support by the University of Basel.
D.L.\ was supported by IARPA, ONR, DARPA, NSA, Sandia National Laboratories and the NIST Quantum Information Program.

\appendix

\section{Summary of used coordinate systems}
\label{app:coordinates}

\newcommand{\coordinates}[1]{\textbf{#1}}
In order to help distinguishing the various coordinate systems used in the text, we summarize them here.
\begin{itemize}
	\item The \coordinates{laboratory frame} is spanned by the unit vectors
		\begin{equation}
			\vec{\hat{x}}=\{1,0,0\},
			\qquad
			\vec{\hat{y}}=\{0,1,0\},
			\qquad
			\vec{\hat{z}}=\{0,0,1\}.
		\end{equation}
		Its orientation is shown in figures~\ref{fig:interaction}, \ref{fig:cpregimes}, and \ref{fig:couplings}. In \sref{sec:cover-plane-boundary} lab-frame vectors are written as $\vec{r}=x\vec{\hat{x}}+y\vec{\hat{y}}+z\vec{\hat{z}}$ with $\rho=\sqrt{x^2+y^2}$ and $r=\sqrt{x^2+y^2+z^2}$.
	\item The \coordinates{pseudo-spin quantization frame} is given by the orthonormal unit vectors $\vec{\hat{X}}$, $\vec{\hat{Y}}$, $\vec{\hat{Z}}$, where $\vec{\hat{Z}}$ is the quantization axis. In \sref{sec:wires} we set $\vec{\hat{Z}}=\vec{\Delta}_Z/d$.
	\item The $i\se{th}$ ion's \coordinates{vibration around its equilibrium position} is expressed in the local coordinate frame $\vec{m}_i^{\mu}$, see \eref{eq:localcoord}. For the Kitaev model we use the vectors given in~\eref{eq:vibrations}: each vibrational direction (depending on which sublattice the ion is located) is indexed by, and associated with, one of the spin-space directions $\vec{\hat{X}}$, $\vec{\hat{Y}}$, $\vec{\hat{Z}}$, but this does not mean that the vibrational directions are parallel (or in any way related) to the spin-space axes.
	\item The vectors \coordinates{connecting neighboring ions} in the Kitaev honeycomb lattice $\vec{\Delta}_X$, $\vec{\Delta}_Y$, $\vec{\Delta}_Z$ are of length $d$ and given in~\eref{eq:latticedirections}. They all lie in the plane of the lattice and do not form a 3D coordinate system.
\end{itemize}

\bibliography{qsim}

\begin{thebibliography}{45}%
\makeatletter
\providecommand \@ifxundefined [1]{%
 \@ifx{#1\undefined}
}%
\providecommand \@ifnum [1]{%
 \ifnum #1\expandafter \@firstoftwo
 \else \expandafter \@secondoftwo
 \fi
}%
\providecommand \@ifx [1]{%
 \ifx #1\expandafter \@firstoftwo
 \else \expandafter \@secondoftwo
 \fi
}%
\providecommand \natexlab [1]{#1}%
\providecommand \enquote  [1]{``#1''}%
\providecommand \bibnamefont  [1]{#1}%
\providecommand \bibfnamefont [1]{#1}%
\providecommand \citenamefont [1]{#1}%
\providecommand \href@noop [0]{\@secondoftwo}%
\providecommand \href [0]{\begingroup \@sanitize@url \@href}%
\providecommand \@href[1]{\@@startlink{#1}\@@href}%
\providecommand \@@href[1]{\endgroup#1\@@endlink}%
\providecommand \@sanitize@url [0]{\catcode `\\12\catcode `\$12\catcode
  `\&12\catcode `\#12\catcode `\^12\catcode `\_12\catcode `\%12\relax}%
\providecommand \@@startlink[1]{}%
\providecommand \@@endlink[0]{}%
\providecommand \url  [0]{\begingroup\@sanitize@url \@url }%
\providecommand \@url [1]{\endgroup\@href {#1}{\urlprefix }}%
\providecommand \urlprefix  [0]{URL }%
\providecommand \Eprint [0]{\href }%
\providecommand \doibase [0]{http://dx.doi.org/}%
\providecommand \selectlanguage [0]{\@gobble}%
\providecommand \bibinfo  [0]{\@secondoftwo}%
\providecommand \bibfield  [0]{\@secondoftwo}%
\providecommand \translation [1]{[#1]}%
\providecommand \BibitemOpen [0]{}%
\providecommand \bibitemStop [0]{}%
\providecommand \bibitemNoStop [0]{.\EOS\space}%
\providecommand \EOS [0]{\spacefactor3000\relax}%
\providecommand \BibitemShut  [1]{\csname bibitem#1\endcsname}%
\let\auto@bib@innerbib\@empty
%</preamble>
\bibitem [{\citenamefont {Cirac}\ and\ \citenamefont
  {Zoller}(1995)}]{Cirac1995}%
  \BibitemOpen
  \bibfield  {author} {\bibinfo {author} {\bibfnamefont {J.~I.}\ \bibnamefont
  {Cirac}}\ and\ \bibinfo {author} {\bibfnamefont {P.}~\bibnamefont {Zoller}},\
  }\href {\doibase 10.1103/PhysRevLett.74.4091} {\bibfield  {journal} {\bibinfo
   {journal} {Phys. Rev. Lett.}\ }\textbf {\bibinfo {volume} {74}},\ \bibinfo
  {pages} {4091} (\bibinfo {year} {1995})}\BibitemShut {NoStop}%
\bibitem [{\citenamefont {S{\o}rensen}\ and\ \citenamefont
  {M{\o}lmer}(1999)}]{Soerensen1999}%
  \BibitemOpen
  \bibfield  {author} {\bibinfo {author} {\bibfnamefont {A.}~\bibnamefont
  {S{\o}rensen}}\ and\ \bibinfo {author} {\bibfnamefont {K.}~\bibnamefont
  {M{\o}lmer}},\ }\href {\doibase 10.1103/PhysRevLett.82.1971} {\bibfield
  {journal} {\bibinfo  {journal} {Phys. Rev. Lett.}\ }\textbf {\bibinfo
  {volume} {82}},\ \bibinfo {pages} {1971} (\bibinfo {year}
  {1999})}\BibitemShut {NoStop}%
\bibitem [{\citenamefont {M{\o}lmer}\ and\ \citenamefont
  {S{\o}rensen}(1999)}]{Moelmer1999a}%
  \BibitemOpen
  \bibfield  {author} {\bibinfo {author} {\bibfnamefont {K.}~\bibnamefont
  {M{\o}lmer}}\ and\ \bibinfo {author} {\bibfnamefont {A.}~\bibnamefont
  {S{\o}rensen}},\ }\href {\doibase 10.1103/PhysRevLett.82.1835} {\bibfield
  {journal} {\bibinfo  {journal} {Phys. Rev. Lett.}\ }\textbf {\bibinfo
  {volume} {82}},\ \bibinfo {pages} {1835} (\bibinfo {year}
  {1999})}\BibitemShut {NoStop}%
\bibitem [{\citenamefont {Mintert}\ and\ \citenamefont
  {Wunderlich}(2001)}]{Mintert2001}%
  \BibitemOpen
  \bibfield  {author} {\bibinfo {author} {\bibfnamefont {F.}~\bibnamefont
  {Mintert}}\ and\ \bibinfo {author} {\bibfnamefont {C.}~\bibnamefont
  {Wunderlich}},\ }\href {\doibase 10.1103/PhysRevLett.87.257904} {\bibfield
  {journal} {\bibinfo  {journal} {Phys. Rev. Lett.}\ }\textbf {\bibinfo
  {volume} {87}},\ \bibinfo {pages} {257904} (\bibinfo {year}
  {2001})}\BibitemShut {NoStop}%
\bibitem [{\citenamefont {Leibfried}\ \emph {et~al.}(2003)\citenamefont
  {Leibfried}, \citenamefont {DeMarco}, \citenamefont {Meyer}, \citenamefont
  {Lucas}, \citenamefont {Barrett}, \citenamefont {Britton}, \citenamefont
  {Itano}, \citenamefont {Jelenkovi\'c}, \citenamefont {Langer}, \citenamefont
  {Rosenband},\ and\ \citenamefont {Wineland}}]{Leibfried2003b}%
  \BibitemOpen
  \bibfield  {author} {\bibinfo {author} {\bibfnamefont {D.}~\bibnamefont
  {Leibfried}}, \bibinfo {author} {\bibfnamefont {B.}~\bibnamefont {DeMarco}},
  \bibinfo {author} {\bibfnamefont {V.}~\bibnamefont {Meyer}}, \bibinfo
  {author} {\bibfnamefont {D.}~\bibnamefont {Lucas}}, \bibinfo {author}
  {\bibfnamefont {M.}~\bibnamefont {Barrett}}, \bibinfo {author} {\bibfnamefont
  {J.}~\bibnamefont {Britton}}, \bibinfo {author} {\bibfnamefont {W.~M.}\
  \bibnamefont {Itano}}, \bibinfo {author} {\bibfnamefont {B.}~\bibnamefont
  {Jelenkovi\'c}}, \bibinfo {author} {\bibfnamefont {C.}~\bibnamefont
  {Langer}}, \bibinfo {author} {\bibfnamefont {T.}~\bibnamefont {Rosenband}}, \
  and\ \bibinfo {author} {\bibfnamefont {D.~J.}\ \bibnamefont {Wineland}},\
  }\href {\doibase 10.1038/nature01492} {\bibfield  {journal} {\bibinfo
  {journal} {Nature}\ }\textbf {\bibinfo {volume} {422}},\ \bibinfo {pages}
  {412} (\bibinfo {year} {2003})}\BibitemShut {NoStop}%
\bibitem [{\citenamefont {Porras}\ and\ \citenamefont
  {Cirac}(2004)}]{Porras2004}%
  \BibitemOpen
  \bibfield  {author} {\bibinfo {author} {\bibfnamefont {D.}~\bibnamefont
  {Porras}}\ and\ \bibinfo {author} {\bibfnamefont {J.~I.}\ \bibnamefont
  {Cirac}},\ }\href {\doibase 10.1103/PhysRevLett.92.207901} {\bibfield
  {journal} {\bibinfo  {journal} {Phys. Rev. Lett.}\ }\textbf {\bibinfo
  {volume} {92}},\ \bibinfo {pages} {207901} (\bibinfo {year}
  {2004})}\BibitemShut {NoStop}%
\bibitem [{\citenamefont {Garc\'ia-Ripoll}\ \emph {et~al.}(2005)\citenamefont
  {Garc\'ia-Ripoll}, \citenamefont {Zoller},\ and\ \citenamefont
  {Cirac}}]{GarciaRipoll2005}%
  \BibitemOpen
  \bibfield  {author} {\bibinfo {author} {\bibfnamefont {J.~J.}\ \bibnamefont
  {Garc\'ia-Ripoll}}, \bibinfo {author} {\bibfnamefont {P.}~\bibnamefont
  {Zoller}}, \ and\ \bibinfo {author} {\bibfnamefont {J.~I.}\ \bibnamefont
  {Cirac}},\ }\href {\doibase 10.1103/PhysRevA.71.062309} {\bibfield  {journal}
  {\bibinfo  {journal} {Phys. Rev. A}\ }\textbf {\bibinfo {volume} {71}},\
  \bibinfo {pages} {062309} (\bibinfo {year} {2005})}\BibitemShut {NoStop}%
\bibitem [{\citenamefont {Porras}\ and\ \citenamefont
  {Cirac}(2006)}]{Porras2006b}%
  \BibitemOpen
  \bibfield  {author} {\bibinfo {author} {\bibfnamefont {D.}~\bibnamefont
  {Porras}}\ and\ \bibinfo {author} {\bibfnamefont {J.~I.}\ \bibnamefont
  {Cirac}},\ }\href {\doibase 10.1103/PhysRevLett.96.250501} {\bibfield
  {journal} {\bibinfo  {journal} {Phys. Rev. Lett.}\ }\textbf {\bibinfo
  {volume} {96}},\ \bibinfo {pages} {250501} (\bibinfo {year}
  {2006})}\BibitemShut {NoStop}%
\bibitem [{\citenamefont {Bermudez}\ \emph {et~al.}(2009)\citenamefont
  {Bermudez}, \citenamefont {Porras},\ and\ \citenamefont
  {Martin-Delgado}}]{Bermudez2009}%
  \BibitemOpen
  \bibfield  {author} {\bibinfo {author} {\bibfnamefont {A.}~\bibnamefont
  {Bermudez}}, \bibinfo {author} {\bibfnamefont {D.}~\bibnamefont {Porras}}, \
  and\ \bibinfo {author} {\bibfnamefont {M.~A.}\ \bibnamefont
  {Martin-Delgado}},\ }\href {\doibase 10.1103/PhysRevA.79.060303} {\bibfield
  {journal} {\bibinfo  {journal} {Phys. Rev. A}\ }\textbf {\bibinfo {volume}
  {79}},\ \bibinfo {pages} {060303(R)} (\bibinfo {year} {2009})}\BibitemShut
  {NoStop}%
\bibitem [{\citenamefont {Welzel}\ \emph {et~al.}(2011)\citenamefont {Welzel},
  \citenamefont {Bautista-Salvador}, \citenamefont {Abarbanel}, \citenamefont
  {Wineman-Fisher}, \citenamefont {Wunderlich}, \citenamefont {Folman},\ and\
  \citenamefont {Schmidt-Kaler}}]{Welzel2011}%
  \BibitemOpen
  \bibfield  {author} {\bibinfo {author} {\bibfnamefont {J.}~\bibnamefont
  {Welzel}}, \bibinfo {author} {\bibfnamefont {A.}~\bibnamefont
  {Bautista-Salvador}}, \bibinfo {author} {\bibfnamefont {C.}~\bibnamefont
  {Abarbanel}}, \bibinfo {author} {\bibfnamefont {V.}~\bibnamefont
  {Wineman-Fisher}}, \bibinfo {author} {\bibfnamefont {C.}~\bibnamefont
  {Wunderlich}}, \bibinfo {author} {\bibfnamefont {R.}~\bibnamefont {Folman}},
  \ and\ \bibinfo {author} {\bibfnamefont {F.}~\bibnamefont {Schmidt-Kaler}},\
  }\href@noop {} {} (\bibinfo {year} {2011}),\ \Eprint
  {http://arxiv.org/abs/1102.3645} {arXiv:1102.3645 [quant-ph]} \BibitemShut
  {NoStop}%
\bibitem [{\citenamefont {Wunderlich}(2002)}]{Wunderlich2002}%
  \BibitemOpen
  \bibfield  {author} {\bibinfo {author} {\bibfnamefont {C.}~\bibnamefont
  {Wunderlich}},\ }\enquote {\bibinfo {title} {Laser physics at the limit},}\ \
  (\bibinfo  {publisher} {Springer},\ \bibinfo {address} {Berlin},\ \bibinfo
  {year} {2002})\ pp.\ \bibinfo {pages} {261--271},\ \Eprint
  {http://arxiv.org/abs/quant-ph/0111158} {arXiv:quant-ph/0111158} \BibitemShut
  {NoStop}%
\bibitem [{\citenamefont {Deng}\ \emph {et~al.}(2005)\citenamefont {Deng},
  \citenamefont {Porras},\ and\ \citenamefont {Cirac}}]{Deng2005}%
  \BibitemOpen
  \bibfield  {author} {\bibinfo {author} {\bibfnamefont {X.-L.}\ \bibnamefont
  {Deng}}, \bibinfo {author} {\bibfnamefont {D.}~\bibnamefont {Porras}}, \ and\
  \bibinfo {author} {\bibfnamefont {J.~I.}\ \bibnamefont {Cirac}},\ }\href
  {\doibase 10.1103/PhysRevA.72.063407} {\bibfield  {journal} {\bibinfo
  {journal} {Phys. Rev. A}\ }\textbf {\bibinfo {volume} {72}},\ \bibinfo
  {pages} {063407} (\bibinfo {year} {2005})}\BibitemShut {NoStop}%
\bibitem [{\citenamefont {Friedenauer}\ \emph {et~al.}(2008)\citenamefont
  {Friedenauer}, \citenamefont {Schmitz}, \citenamefont {Glueckert},
  \citenamefont {Porras},\ and\ \citenamefont {Schaetz}}]{Friedenauer2008}%
  \BibitemOpen
  \bibfield  {author} {\bibinfo {author} {\bibfnamefont {A.}~\bibnamefont
  {Friedenauer}}, \bibinfo {author} {\bibfnamefont {H.}~\bibnamefont
  {Schmitz}}, \bibinfo {author} {\bibfnamefont {J.~T.}\ \bibnamefont
  {Glueckert}}, \bibinfo {author} {\bibfnamefont {D.}~\bibnamefont {Porras}}, \
  and\ \bibinfo {author} {\bibfnamefont {T.}~\bibnamefont {Schaetz}},\ }\href
  {\doibase 10.1038/nphys1032} {\bibfield  {journal} {\bibinfo  {journal}
  {Nature Physics}\ }\textbf {\bibinfo {volume} {4}},\ \bibinfo {pages} {757}
  (\bibinfo {year} {2008})}\BibitemShut {NoStop}%
\bibitem [{\citenamefont {Johanning}\ \emph
  {et~al.}(2009{\natexlab{a}})\citenamefont {Johanning}, \citenamefont
  {Var{\'o}n},\ and\ \citenamefont {Wunderlich}}]{Johanning2009}%
  \BibitemOpen
  \bibfield  {author} {\bibinfo {author} {\bibfnamefont {M.}~\bibnamefont
  {Johanning}}, \bibinfo {author} {\bibfnamefont {A.~F.}\ \bibnamefont
  {Var{\'o}n}}, \ and\ \bibinfo {author} {\bibfnamefont {C.}~\bibnamefont
  {Wunderlich}},\ }\href {\doibase 10.1088/0953-4075/42/15/154009} {\bibfield
  {journal} {\bibinfo  {journal} {J. Phys. B: At. Mol. Opt. Phys.}\ }\textbf
  {\bibinfo {volume} {42}},\ \bibinfo {pages} {154009} (\bibinfo {year}
  {2009}{\natexlab{a}})}\BibitemShut {NoStop}%
\bibitem [{\citenamefont {Kim}\ \emph {et~al.}(2010)\citenamefont {Kim},
  \citenamefont {Chang}, \citenamefont {Korenblit}, \citenamefont {Islam},
  \citenamefont {Edwards}, \citenamefont {Freericks}, \citenamefont {Lin},
  \citenamefont {Duan},\ and\ \citenamefont {Monroe}}]{Kim2010}%
  \BibitemOpen
  \bibfield  {author} {\bibinfo {author} {\bibfnamefont {K.}~\bibnamefont
  {Kim}}, \bibinfo {author} {\bibfnamefont {M.-S.}\ \bibnamefont {Chang}},
  \bibinfo {author} {\bibfnamefont {S.}~\bibnamefont {Korenblit}}, \bibinfo
  {author} {\bibfnamefont {R.}~\bibnamefont {Islam}}, \bibinfo {author}
  {\bibfnamefont {E.~E.}\ \bibnamefont {Edwards}}, \bibinfo {author}
  {\bibfnamefont {J.~K.}\ \bibnamefont {Freericks}}, \bibinfo {author}
  {\bibfnamefont {G.-D.}\ \bibnamefont {Lin}}, \bibinfo {author} {\bibfnamefont
  {L.-M.}\ \bibnamefont {Duan}}, \ and\ \bibinfo {author} {\bibfnamefont
  {C.}~\bibnamefont {Monroe}},\ }\href {\doibase 10.1038/nature09071}
  {\bibfield  {journal} {\bibinfo  {journal} {Nature}\ }\textbf {\bibinfo
  {volume} {465}},\ \bibinfo {pages} {590} (\bibinfo {year}
  {2010})}\BibitemShut {NoStop}%
\bibitem [{\citenamefont {S{\o}rensen}\ and\ \citenamefont
  {M{\o}lmer}(2000)}]{soerensen00:entan_quant_comput_with_ions}%
  \BibitemOpen
  \bibfield  {author} {\bibinfo {author} {\bibfnamefont {A.}~\bibnamefont
  {S{\o}rensen}}\ and\ \bibinfo {author} {\bibfnamefont {K.}~\bibnamefont
  {M{\o}lmer}},\ }\href {\doibase 10.1103/PhysRevA.62.022311} {\bibfield
  {journal} {\bibinfo  {journal} {Phys. Rev. A}\ }\textbf {\bibinfo {volume}
  {62}},\ \bibinfo {pages} {022311} (\bibinfo {year} {2000})}\BibitemShut
  {NoStop}%
\bibitem [{\citenamefont {Milburn}\ \emph {et~al.}(2000)\citenamefont
  {Milburn}, \citenamefont {Schneider},\ and\ \citenamefont
  {James}}]{milburn00:ion_trap_quant_comput_with}%
  \BibitemOpen
  \bibfield  {author} {\bibinfo {author} {\bibfnamefont {G.~J.}\ \bibnamefont
  {Milburn}}, \bibinfo {author} {\bibfnamefont {S.}~\bibnamefont {Schneider}},
  \ and\ \bibinfo {author} {\bibfnamefont {D.~F.}\ \bibnamefont {James}},\
  }\href {\doibase
  10.1002/1521-3978(200009)48:9/11<801::AID-PROP801>3.0.CO;2-1} {\bibfield
  {journal} {\bibinfo  {journal} {Fortschr. Phys.}\ }\textbf {\bibinfo {volume}
  {48}},\ \bibinfo {pages} {801} (\bibinfo {year} {2000})}\BibitemShut
  {NoStop}%
\bibitem [{\citenamefont {Lee}\ \emph {et~al.}(2005)\citenamefont {Lee},
  \citenamefont {Brickman}, \citenamefont {Deslauriers}, \citenamefont
  {Haljan}, \citenamefont {Duan},\ and\ \citenamefont {Monroe}}]{Lee2005}%
  \BibitemOpen
  \bibfield  {author} {\bibinfo {author} {\bibfnamefont {P.~J.}\ \bibnamefont
  {Lee}}, \bibinfo {author} {\bibfnamefont {K.-A.}\ \bibnamefont {Brickman}},
  \bibinfo {author} {\bibfnamefont {L.}~\bibnamefont {Deslauriers}}, \bibinfo
  {author} {\bibfnamefont {P.~C.}\ \bibnamefont {Haljan}}, \bibinfo {author}
  {\bibfnamefont {L.-M.}\ \bibnamefont {Duan}}, \ and\ \bibinfo {author}
  {\bibfnamefont {C.}~\bibnamefont {Monroe}},\ }\href {\doibase
  10.1088/1464-4266/7/10/025} {\bibfield  {journal} {\bibinfo  {journal} {J.
  Opt. B: Quantum Semiclass. Opt.}\ }\textbf {\bibinfo {volume} {7}},\ \bibinfo
  {pages} {S371} (\bibinfo {year} {2005})}\BibitemShut {NoStop}%
\bibitem [{\citenamefont {Kim}\ \emph {et~al.}(2009)\citenamefont {Kim},
  \citenamefont {Chang}, \citenamefont {Islam}, \citenamefont {Korenblit},
  \citenamefont {Duan},\ and\ \citenamefont {Monroe}}]{Kim2009}%
  \BibitemOpen
  \bibfield  {author} {\bibinfo {author} {\bibfnamefont {K.}~\bibnamefont
  {Kim}}, \bibinfo {author} {\bibfnamefont {M.-S.}\ \bibnamefont {Chang}},
  \bibinfo {author} {\bibfnamefont {R.}~\bibnamefont {Islam}}, \bibinfo
  {author} {\bibfnamefont {S.}~\bibnamefont {Korenblit}}, \bibinfo {author}
  {\bibfnamefont {L.-M.}\ \bibnamefont {Duan}}, \ and\ \bibinfo {author}
  {\bibfnamefont {C.}~\bibnamefont {Monroe}},\ }\href {\doibase
  10.1103/PhysRevLett.103.120502} {\bibfield  {journal} {\bibinfo  {journal}
  {Phys. Rev. Lett.}\ }\textbf {\bibinfo {volume} {103}},\ \bibinfo {pages}
  {120502} (\bibinfo {year} {2009})}\BibitemShut {NoStop}%
\bibitem [{\citenamefont {Johanning}\ \emph
  {et~al.}(2009{\natexlab{b}})\citenamefont {Johanning}, \citenamefont {Braun},
  \citenamefont {Timoney}, \citenamefont {Elman}, \citenamefont {Neuhauser},\
  and\ \citenamefont {Wunderlich}}]{Johanning2009b}%
  \BibitemOpen
  \bibfield  {author} {\bibinfo {author} {\bibfnamefont {M.}~\bibnamefont
  {Johanning}}, \bibinfo {author} {\bibfnamefont {A.}~\bibnamefont {Braun}},
  \bibinfo {author} {\bibfnamefont {N.}~\bibnamefont {Timoney}}, \bibinfo
  {author} {\bibfnamefont {V.}~\bibnamefont {Elman}}, \bibinfo {author}
  {\bibfnamefont {W.}~\bibnamefont {Neuhauser}}, \ and\ \bibinfo {author}
  {\bibfnamefont {C.}~\bibnamefont {Wunderlich}},\ }\href {\doibase
  10.1103/PhysRevLett.102.073004} {\bibfield  {journal} {\bibinfo  {journal}
  {Phys. Rev. Lett.}\ }\textbf {\bibinfo {volume} {102}},\ \bibinfo {pages}
  {073004} (\bibinfo {year} {2009}{\natexlab{b}})}\BibitemShut {NoStop}%
\bibitem [{\citenamefont {Ospelkaus}\ \emph {et~al.}(2008)\citenamefont
  {Ospelkaus}, \citenamefont {Langer}, \citenamefont {Amini}, \citenamefont
  {Brown}, \citenamefont {Leibfried},\ and\ \citenamefont
  {Wineland}}]{Ospelkaus2008}%
  \BibitemOpen
  \bibfield  {author} {\bibinfo {author} {\bibfnamefont {C.}~\bibnamefont
  {Ospelkaus}}, \bibinfo {author} {\bibfnamefont {C.~E.}\ \bibnamefont
  {Langer}}, \bibinfo {author} {\bibfnamefont {J.~M.}\ \bibnamefont {Amini}},
  \bibinfo {author} {\bibfnamefont {K.~R.}\ \bibnamefont {Brown}}, \bibinfo
  {author} {\bibfnamefont {D.}~\bibnamefont {Leibfried}}, \ and\ \bibinfo
  {author} {\bibfnamefont {D.~J.}\ \bibnamefont {Wineland}},\ }\href {\doibase
  10.1103/PhysRevLett.101.090502} {\bibfield  {journal} {\bibinfo  {journal}
  {Phys. Rev. Lett.}\ }\textbf {\bibinfo {volume} {101}},\ \bibinfo {pages}
  {090502} (\bibinfo {year} {2008})}\BibitemShut {NoStop}%
\bibitem [{\citenamefont {Ospelkaus}\ \emph {et~al.}(2011)\citenamefont
  {Ospelkaus}, \citenamefont {Warring}, \citenamefont {Colombe}, \citenamefont
  {Brown}, \citenamefont {Amini}, \citenamefont {Leibfried},\ and\
  \citenamefont {Wineland}}]{Ospelkaus2011}%
  \BibitemOpen
  \bibfield  {author} {\bibinfo {author} {\bibfnamefont {C.}~\bibnamefont
  {Ospelkaus}}, \bibinfo {author} {\bibfnamefont {U.}~\bibnamefont {Warring}},
  \bibinfo {author} {\bibfnamefont {Y.}~\bibnamefont {Colombe}}, \bibinfo
  {author} {\bibfnamefont {K.~R.}\ \bibnamefont {Brown}}, \bibinfo {author}
  {\bibfnamefont {J.~M.}\ \bibnamefont {Amini}}, \bibinfo {author}
  {\bibfnamefont {D.}~\bibnamefont {Leibfried}}, \ and\ \bibinfo {author}
  {\bibfnamefont {D.~J.}\ \bibnamefont {Wineland}},\ }\href {\doibase
  10.1038/nature10290} {\bibfield  {journal} {\bibinfo  {journal} {Nature}\
  }\textbf {\bibinfo {volume} {476}},\ \bibinfo {pages} {181} (\bibinfo {year}
  {2011})}\BibitemShut {NoStop}%
\bibitem [{\citenamefont {Brown}\ \emph {et~al.}(2011)\citenamefont {Brown},
  \citenamefont {Ospelkaus}, \citenamefont {Colombe}, \citenamefont {Wilson},
  \citenamefont {Leibfried},\ and\ \citenamefont
  {Wineland}}]{Brown11:Coupled-quantiz}%
  \BibitemOpen
  \bibfield  {author} {\bibinfo {author} {\bibfnamefont {K.~R.}\ \bibnamefont
  {Brown}}, \bibinfo {author} {\bibfnamefont {C.}~\bibnamefont {Ospelkaus}},
  \bibinfo {author} {\bibfnamefont {Y.}~\bibnamefont {Colombe}}, \bibinfo
  {author} {\bibfnamefont {A.~C.}\ \bibnamefont {Wilson}}, \bibinfo {author}
  {\bibfnamefont {D.}~\bibnamefont {Leibfried}}, \ and\ \bibinfo {author}
  {\bibfnamefont {D.~J.}\ \bibnamefont {Wineland}},\ }\href {\doibase
  10.1038/nature09721} {\bibfield  {journal} {\bibinfo  {journal} {Nature}\
  }\textbf {\bibinfo {volume} {471}},\ \bibinfo {pages} {196} (\bibinfo {year}
  {2011})}\BibitemShut {NoStop}%
\bibitem [{\citenamefont {Harlander}\ \emph {et~al.}(2011)\citenamefont
  {Harlander}, \citenamefont {Lechner}, \citenamefont {Brownnutt},
  \citenamefont {Blatt},\ and\ \citenamefont
  {H{\"a}nsel}}]{Harlander11:Trapped-ion-ant}%
  \BibitemOpen
  \bibfield  {author} {\bibinfo {author} {\bibfnamefont {M.}~\bibnamefont
  {Harlander}}, \bibinfo {author} {\bibfnamefont {R.}~\bibnamefont {Lechner}},
  \bibinfo {author} {\bibfnamefont {M.}~\bibnamefont {Brownnutt}}, \bibinfo
  {author} {\bibfnamefont {R.}~\bibnamefont {Blatt}}, \ and\ \bibinfo {author}
  {\bibfnamefont {W.}~\bibnamefont {H{\"a}nsel}},\ }\href {\doibase
  10.1038/nature09800} {\bibfield  {journal} {\bibinfo  {journal} {Nature}\
  }\textbf {\bibinfo {volume} {471}},\ \bibinfo {pages} {200} (\bibinfo {year}
  {2011})}\BibitemShut {NoStop}%
\bibitem [{\citenamefont {Kitaev}(2006)}]{Kitaev2006}%
  \BibitemOpen
  \bibfield  {author} {\bibinfo {author} {\bibfnamefont {A.}~\bibnamefont
  {Kitaev}},\ }\href {\doibase 10.1016/j.aop.2005.10.005} {\bibfield  {journal}
  {\bibinfo  {journal} {Annals of Physics}\ }\textbf {\bibinfo {volume}
  {321}},\ \bibinfo {pages} {2} (\bibinfo {year} {2006})}\BibitemShut {NoStop}%
\bibitem [{\citenamefont {Taddei}\ \emph {et~al.}(2009)\citenamefont {Taddei},
  \citenamefont {Mendes},\ and\ \citenamefont {Farina}}]{Taddei2009}%
  \BibitemOpen
  \bibfield  {author} {\bibinfo {author} {\bibfnamefont {M.~M.}\ \bibnamefont
  {Taddei}}, \bibinfo {author} {\bibfnamefont {T.~N.~C.}\ \bibnamefont
  {Mendes}}, \ and\ \bibinfo {author} {\bibfnamefont {C.}~\bibnamefont
  {Farina}},\ }\href {\doibase 10.1088/0143-0807/30/5/005} {\bibfield
  {journal} {\bibinfo  {journal} {Eur. J. Phys.}\ }\textbf {\bibinfo {volume}
  {30}},\ \bibinfo {pages} {965} (\bibinfo {year} {2009})}\BibitemShut
  {NoStop}%
\bibitem [{\citenamefont {Jackson}(1999)}]{Jackson}%
  \BibitemOpen
  \bibfield  {author} {\bibinfo {author} {\bibfnamefont {J.~D.}\ \bibnamefont
  {Jackson}},\ }\href@noop {} {\emph {\bibinfo {title} {Classical
  Electrodynamics}}},\ \bibinfo {edition} {3rd}\ ed.\ (\bibinfo  {publisher}
  {John Wiley \& Sons},\ \bibinfo {year} {1999})\BibitemShut {NoStop}%
\bibitem [{\citenamefont {Chiaverini}\ \emph {et~al.}(2005)\citenamefont
  {Chiaverini}, \citenamefont {Blakestad}, \citenamefont {Britton},
  \citenamefont {Jost}, \citenamefont {Langer}, \citenamefont {Leibfried},
  \citenamefont {Ozeri},\ and\ \citenamefont {Wineland}}]{Chiaverini2005}%
  \BibitemOpen
  \bibfield  {author} {\bibinfo {author} {\bibfnamefont {J.}~\bibnamefont
  {Chiaverini}}, \bibinfo {author} {\bibfnamefont {R.~B.}\ \bibnamefont
  {Blakestad}}, \bibinfo {author} {\bibfnamefont {J.}~\bibnamefont {Britton}},
  \bibinfo {author} {\bibfnamefont {J.~D.}\ \bibnamefont {Jost}}, \bibinfo
  {author} {\bibfnamefont {C.}~\bibnamefont {Langer}}, \bibinfo {author}
  {\bibfnamefont {D.}~\bibnamefont {Leibfried}}, \bibinfo {author}
  {\bibfnamefont {R.}~\bibnamefont {Ozeri}}, \ and\ \bibinfo {author}
  {\bibfnamefont {D.~J.}\ \bibnamefont {Wineland}},\ }\href@noop {} {\bibfield
  {journal} {\bibinfo  {journal} {Quant. Inf. and Comp.}\ }\textbf {\bibinfo
  {volume} {5}},\ \bibinfo {pages} {419} (\bibinfo {year} {2005})}\BibitemShut
  {NoStop}%
\bibitem [{\citenamefont {Seidelin}\ \emph {et~al.}(2006)\citenamefont
  {Seidelin}, \citenamefont {Chiaverini}, \citenamefont {Reichle},
  \citenamefont {Bollinger}, \citenamefont {Leibfried}, \citenamefont
  {Britton}, \citenamefont {Wesenberg}, \citenamefont {Blakestad},
  \citenamefont {Epstein}, \citenamefont {Hume}, \citenamefont {Itano},
  \citenamefont {Jost}, \citenamefont {Langer}, \citenamefont {Ozeri},
  \citenamefont {Shiga},\ and\ \citenamefont {Wineland}}]{Seidelin2006}%
  \BibitemOpen
  \bibfield  {author} {\bibinfo {author} {\bibfnamefont {S.}~\bibnamefont
  {Seidelin}}, \bibinfo {author} {\bibfnamefont {J.}~\bibnamefont
  {Chiaverini}}, \bibinfo {author} {\bibfnamefont {R.}~\bibnamefont {Reichle}},
  \bibinfo {author} {\bibfnamefont {J.~J.}\ \bibnamefont {Bollinger}}, \bibinfo
  {author} {\bibfnamefont {D.}~\bibnamefont {Leibfried}}, \bibinfo {author}
  {\bibfnamefont {J.}~\bibnamefont {Britton}}, \bibinfo {author} {\bibfnamefont
  {J.~H.}\ \bibnamefont {Wesenberg}}, \bibinfo {author} {\bibfnamefont {R.~B.}\
  \bibnamefont {Blakestad}}, \bibinfo {author} {\bibfnamefont {R.~J.}\
  \bibnamefont {Epstein}}, \bibinfo {author} {\bibfnamefont {D.~B.}\
  \bibnamefont {Hume}}, \bibinfo {author} {\bibfnamefont {W.~M.}\ \bibnamefont
  {Itano}}, \bibinfo {author} {\bibfnamefont {J.~D.}\ \bibnamefont {Jost}},
  \bibinfo {author} {\bibfnamefont {C.}~\bibnamefont {Langer}}, \bibinfo
  {author} {\bibfnamefont {R.}~\bibnamefont {Ozeri}}, \bibinfo {author}
  {\bibfnamefont {N.}~\bibnamefont {Shiga}}, \ and\ \bibinfo {author}
  {\bibfnamefont {D.~J.}\ \bibnamefont {Wineland}},\ }\href {\doibase
  10.1103/PhysRevLett.96.253003} {\bibfield  {journal} {\bibinfo  {journal}
  {Phys. Rev. Lett.}\ }\textbf {\bibinfo {volume} {96}},\ \bibinfo {pages}
  {253003} (\bibinfo {year} {2006})}\BibitemShut {NoStop}%
\bibitem [{\citenamefont {Pearson}\ \emph {et~al.}(2006)\citenamefont
  {Pearson}, \citenamefont {Leibrand}, \citenamefont {Bakr}, \citenamefont
  {Mallard}, \citenamefont {Brown},\ and\ \citenamefont
  {Chuang}}]{Pearson2006}%
  \BibitemOpen
  \bibfield  {author} {\bibinfo {author} {\bibfnamefont {C.~E.}\ \bibnamefont
  {Pearson}}, \bibinfo {author} {\bibfnamefont {D.~R.}\ \bibnamefont
  {Leibrand}}, \bibinfo {author} {\bibfnamefont {W.~S.}\ \bibnamefont {Bakr}},
  \bibinfo {author} {\bibfnamefont {W.~J.}\ \bibnamefont {Mallard}}, \bibinfo
  {author} {\bibfnamefont {K.~R.}\ \bibnamefont {Brown}}, \ and\ \bibinfo
  {author} {\bibfnamefont {I.~L.}\ \bibnamefont {Chuang}},\ }\href {\doibase
  10.1103/PhysRevA.73.032307} {\bibfield  {journal} {\bibinfo  {journal} {Phys.
  Rev. A}\ }\textbf {\bibinfo {volume} {73}},\ \bibinfo {pages} {032307}
  (\bibinfo {year} {2006})}\BibitemShut {NoStop}%
\bibitem [{\citenamefont {VanDevender}\ \emph {et~al.}(2010)\citenamefont
  {VanDevender}, \citenamefont {Colombe}, \citenamefont {Amini}, \citenamefont
  {Leibfried},\ and\ \citenamefont {Wineland}}]{VanDevender2010}%
  \BibitemOpen
  \bibfield  {author} {\bibinfo {author} {\bibfnamefont {A.~P.}\ \bibnamefont
  {VanDevender}}, \bibinfo {author} {\bibfnamefont {Y.}~\bibnamefont
  {Colombe}}, \bibinfo {author} {\bibfnamefont {J.}~\bibnamefont {Amini}},
  \bibinfo {author} {\bibfnamefont {D.}~\bibnamefont {Leibfried}}, \ and\
  \bibinfo {author} {\bibfnamefont {D.~J.}\ \bibnamefont {Wineland}},\ }\href
  {\doibase 10.1103/PhysRevLett.105.023001} {\bibfield  {journal} {\bibinfo
  {journal} {Phys. Rev. Lett.}\ }\textbf {\bibinfo {volume} {105}},\ \bibinfo
  {pages} {023001} (\bibinfo {year} {2010})}\BibitemShut {NoStop}%
\bibitem [{\citenamefont {Schmied}(2010)}]{Schmied2010}%
  \BibitemOpen
  \bibfield  {author} {\bibinfo {author} {\bibfnamefont {R.}~\bibnamefont
  {Schmied}},\ }\href {\doibase 10.1088/1367-2630/12/2/023038} {\bibfield
  {journal} {\bibinfo  {journal} {New J. Phys.}\ }\textbf {\bibinfo {volume}
  {12}},\ \bibinfo {pages} {023038} (\bibinfo {year} {2010})}\BibitemShut
  {NoStop}%
\bibitem [{\citenamefont {Oliveira}\ and\ \citenamefont
  {Miranda}(2001)}]{Oliveira2001}%
  \BibitemOpen
  \bibfield  {author} {\bibinfo {author} {\bibfnamefont {M.~H.}\ \bibnamefont
  {Oliveira}}\ and\ \bibinfo {author} {\bibfnamefont {J.~A.}\ \bibnamefont
  {Miranda}},\ }\href {\doibase 10.1088/0143-0807/22/1/304} {\bibfield
  {journal} {\bibinfo  {journal} {Eur. J. Phys.}\ }\textbf {\bibinfo {volume}
  {22}},\ \bibinfo {pages} {31} (\bibinfo {year} {2001})}\BibitemShut {NoStop}%
\bibitem [{\citenamefont {Schmied}\ \emph {et~al.}(2010)\citenamefont
  {Schmied}, \citenamefont {Leibfried}, \citenamefont {Spreeuw},\ and\
  \citenamefont {Whitlock}}]{Schmied2010b}%
  \BibitemOpen
  \bibfield  {author} {\bibinfo {author} {\bibfnamefont {R.}~\bibnamefont
  {Schmied}}, \bibinfo {author} {\bibfnamefont {D.}~\bibnamefont {Leibfried}},
  \bibinfo {author} {\bibfnamefont {R.~J.~C.}\ \bibnamefont {Spreeuw}}, \ and\
  \bibinfo {author} {\bibfnamefont {S.}~\bibnamefont {Whitlock}},\ }\href
  {\doibase 10.1088/1367-2630/12/10/103029} {\bibfield  {journal} {\bibinfo
  {journal} {New J. Phys.}\ }\textbf {\bibinfo {volume} {12}},\ \bibinfo
  {pages} {103029} (\bibinfo {year} {2010})}\BibitemShut {NoStop}%
\bibitem [{\citenamefont {Wesenberg}(2008)}]{Wesenberg2008}%
  \BibitemOpen
  \bibfield  {author} {\bibinfo {author} {\bibfnamefont {J.~H.}\ \bibnamefont
  {Wesenberg}},\ }\href {\doibase 10.1103/PhysRevA.78.063410} {\bibfield
  {journal} {\bibinfo  {journal} {Phys. Rev. A}\ }\textbf {\bibinfo {volume}
  {78}},\ \bibinfo {pages} {063410} (\bibinfo {year} {2008})}\BibitemShut
  {NoStop}%
\bibitem [{\citenamefont {Schmied}\ \emph {et~al.}(2009)\citenamefont
  {Schmied}, \citenamefont {Wesenberg},\ and\ \citenamefont
  {Leibfried}}]{Schmied2009}%
  \BibitemOpen
  \bibfield  {author} {\bibinfo {author} {\bibfnamefont {R.}~\bibnamefont
  {Schmied}}, \bibinfo {author} {\bibfnamefont {J.~H.}\ \bibnamefont
  {Wesenberg}}, \ and\ \bibinfo {author} {\bibfnamefont {D.}~\bibnamefont
  {Leibfried}},\ }\href {\doibase 10.1103/PhysRevLett.102.233002} {\bibfield
  {journal} {\bibinfo  {journal} {Phys. Rev. Lett.}\ }\textbf {\bibinfo
  {volume} {102}},\ \bibinfo {pages} {233002} (\bibinfo {year}
  {2009})}\BibitemShut {NoStop}%
\bibitem [{\citenamefont {Lloyd}(1996)}]{Lloyd1996}%
  \BibitemOpen
  \bibfield  {author} {\bibinfo {author} {\bibfnamefont {S.}~\bibnamefont
  {Lloyd}},\ }\href {\doibase 10.1126/science.273.5278.1073} {\bibfield
  {journal} {\bibinfo  {journal} {Science}\ }\textbf {\bibinfo {volume}
  {273}},\ \bibinfo {pages} {1073} (\bibinfo {year} {1996})}\BibitemShut
  {NoStop}%
\bibitem [{\citenamefont {Goldstein}(1980)}]{gol80}%
  \BibitemOpen
  \bibfield  {author} {\bibinfo {author} {\bibfnamefont {H.}~\bibnamefont
  {Goldstein}},\ }\href@noop {} {\emph {\bibinfo {title} {Classical
  Mechanics}}},\ \bibinfo {edition} {2nd}\ ed.\ (\bibinfo  {publisher}
  {Addison-Wesley},\ \bibinfo {year} {1980})\BibitemShut {NoStop}%
\bibitem [{\citenamefont {Roos}(2008)}]{Roos2008}%
  \BibitemOpen
  \bibfield  {author} {\bibinfo {author} {\bibfnamefont {C.~F.}\ \bibnamefont
  {Roos}},\ }\href {\doibase 10.1088/1367-2630/10/1/013002} {\bibfield
  {journal} {\bibinfo  {journal} {New J. Phys.}\ }\textbf {\bibinfo {volume}
  {10}},\ \bibinfo {pages} {013002} (\bibinfo {year} {2008})}\BibitemShut
  {NoStop}%
\bibitem [{\citenamefont {Schneider}\ \emph {et~al.}(2011)\citenamefont
  {Schneider}, \citenamefont {Porras},\ and\ \citenamefont
  {Schaetz}}]{Schneider2011}%
  \BibitemOpen
  \bibfield  {author} {\bibinfo {author} {\bibfnamefont {C.}~\bibnamefont
  {Schneider}}, \bibinfo {author} {\bibfnamefont {D.}~\bibnamefont {Porras}}, \
  and\ \bibinfo {author} {\bibfnamefont {T.}~\bibnamefont {Schaetz}},\
  }\href@noop {} {} (\bibinfo {year} {2011}),\ \Eprint
  {http://arxiv.org/abs/1106.2597} {arXiv:1106.2597 [quant-ph]} \BibitemShut
  {NoStop}%
\bibitem [{\citenamefont {Nussinov}\ and\ \citenamefont
  {Ortiz}(2008)}]{Nussinov2008}%
  \BibitemOpen
  \bibfield  {author} {\bibinfo {author} {\bibfnamefont {Z.}~\bibnamefont
  {Nussinov}}\ and\ \bibinfo {author} {\bibfnamefont {G.}~\bibnamefont
  {Ortiz}},\ }\href {\doibase 10.1103/PhysRevB.77.064302} {\bibfield  {journal}
  {\bibinfo  {journal} {Phys. Rev. B}\ }\textbf {\bibinfo {volume} {77}},\
  \bibinfo {pages} {064302} (\bibinfo {year} {2008})}\BibitemShut {NoStop}%
\bibitem [{\citenamefont {Hastings}(2011)}]{Hastings2011}%
  \BibitemOpen
  \bibfield  {author} {\bibinfo {author} {\bibfnamefont {M.~B.}\ \bibnamefont
  {Hastings}},\ }\href@noop {} {} (\bibinfo {year} {2011}),\ \Eprint
  {http://arxiv.org/abs/1106.6026} {arXiv:1106.6026 [quant-ph]} \BibitemShut
  {NoStop}%
\bibitem [{\citenamefont {Berkeland}\ \emph {et~al.}(1998)\citenamefont
  {Berkeland}, \citenamefont {Miller}, \citenamefont {Bergquist}, \citenamefont
  {Itano},\ and\ \citenamefont {Wineland}}]{Berkeland1998}%
  \BibitemOpen
  \bibfield  {author} {\bibinfo {author} {\bibfnamefont {D.~J.}\ \bibnamefont
  {Berkeland}}, \bibinfo {author} {\bibfnamefont {J.~D.}\ \bibnamefont
  {Miller}}, \bibinfo {author} {\bibfnamefont {J.~C.}\ \bibnamefont
  {Bergquist}}, \bibinfo {author} {\bibfnamefont {W.~M.}\ \bibnamefont
  {Itano}}, \ and\ \bibinfo {author} {\bibfnamefont {D.~J.}\ \bibnamefont
  {Wineland}},\ }\href {\doibase 10.1063/1.367318} {\bibfield  {journal}
  {\bibinfo  {journal} {J. Appl. Phys.}\ }\textbf {\bibinfo {volume} {83}},\
  \bibinfo {pages} {5025} (\bibinfo {year} {1998})}\BibitemShut {NoStop}%
\bibitem [{\citenamefont {Turchette}\ \emph {et~al.}(2000)\citenamefont
  {Turchette}, \citenamefont {Kielpinski}, \citenamefont {King}, \citenamefont
  {Leibfried}, \citenamefont {Meekhof}, \citenamefont {Myatt}, \citenamefont
  {Rowe}, \citenamefont {Sackett}, \citenamefont {Wood}, \citenamefont {Itano},
  \citenamefont {Monroe},\ and\ \citenamefont {Wineland}}]{Turchette2000}%
  \BibitemOpen
  \bibfield  {author} {\bibinfo {author} {\bibfnamefont {Q.~A.}\ \bibnamefont
  {Turchette}}, \bibinfo {author} {\bibfnamefont {D.}~\bibnamefont
  {Kielpinski}}, \bibinfo {author} {\bibfnamefont {B.~E.}\ \bibnamefont
  {King}}, \bibinfo {author} {\bibfnamefont {D.}~\bibnamefont {Leibfried}},
  \bibinfo {author} {\bibfnamefont {D.~M.}\ \bibnamefont {Meekhof}}, \bibinfo
  {author} {\bibfnamefont {C.~J.}\ \bibnamefont {Myatt}}, \bibinfo {author}
  {\bibfnamefont {M.~A.}\ \bibnamefont {Rowe}}, \bibinfo {author}
  {\bibfnamefont {C.~A.}\ \bibnamefont {Sackett}}, \bibinfo {author}
  {\bibfnamefont {C.~S.}\ \bibnamefont {Wood}}, \bibinfo {author}
  {\bibfnamefont {W.~M.}\ \bibnamefont {Itano}}, \bibinfo {author}
  {\bibfnamefont {C.}~\bibnamefont {Monroe}}, \ and\ \bibinfo {author}
  {\bibfnamefont {D.~J.}\ \bibnamefont {Wineland}},\ }\href {\doibase
  10.1103/PhysRevA.61.063418} {\bibfield  {journal} {\bibinfo  {journal} {Phys.
  Rev. A}\ }\textbf {\bibinfo {volume} {61}},\ \bibinfo {pages} {063418}
  (\bibinfo {year} {2000})}\BibitemShut {NoStop}%
\bibitem [{\citenamefont {Deslauriers}\ \emph {et~al.}(2006)\citenamefont
  {Deslauriers}, \citenamefont {Olmschenk}, \citenamefont {Stick},
  \citenamefont {Hensinger}, \citenamefont {Sterk},\ and\ \citenamefont
  {Monroe}}]{Deslauriers2006}%
  \BibitemOpen
  \bibfield  {author} {\bibinfo {author} {\bibfnamefont {L.}~\bibnamefont
  {Deslauriers}}, \bibinfo {author} {\bibfnamefont {S.}~\bibnamefont
  {Olmschenk}}, \bibinfo {author} {\bibfnamefont {D.}~\bibnamefont {Stick}},
  \bibinfo {author} {\bibfnamefont {W.~K.}\ \bibnamefont {Hensinger}}, \bibinfo
  {author} {\bibfnamefont {J.}~\bibnamefont {Sterk}}, \ and\ \bibinfo {author}
  {\bibfnamefont {C.}~\bibnamefont {Monroe}},\ }\href {\doibase
  10.1103/PhysRevLett.97.103007} {\bibfield  {journal} {\bibinfo  {journal}
  {Phys. Rev. Lett.}\ }\textbf {\bibinfo {volume} {97}},\ \bibinfo {pages}
  {103007} (\bibinfo {year} {2006})}\BibitemShut {NoStop}%
\end{thebibliography}%
\end{document}